\documentclass[twocolumn]{aastex63}

\newcommand{\Msun}{\ensuremath{{\rm M}_{\odot}}}
\newcommand{\f}{\frac}
\newcommand{\der}{\ensuremath{{\rm d}}}

\shorttitle{Extending the EoR window}
\shortauthors{Raut, Dinesh}

\usepackage{mathptmx}
\usepackage{mathtools}

\begin{document}
\title{Extending the Epoch of Reionization window with apt Foreground and Instrument modeling}

\correspondingauthor{Raut Dinesh}
\email{dinesh.v.raut@gmail.com}

\author{Raut Dinesh}
\affiliation{1079 Shukrawar Peth, Pune 411002, India}

\begin{abstract}
It is seen that foregrounds of 21cm Epoch of Reionization experiments, which are
expected to have smooth spectral dependence, are dominant in a wedge shaped region
of the Fourier space called as Foreground Wedge. A possible way forward to isolate the
21cm Epoch of Reionization (EoR) signal from the much larger foreground component
is to focus on the remaining portion of the Fourier space called as the EoR window. There
are in-fact three distinct regions in the Fourier space, (i) the Foreground wedge portion,
(ii) the EoR window region which lies beyond Horizon wedge and (iii) the region in
between the Horizon wedge and the Field of View wedge. 
This paper addresses two questions: 
(1) Whether the signal in between the two wedges is also a direct representation of the EoR brightness temperature fluctuations?
and (2) How can we extract the cosmology information from this region considering that the foregrounds are much larger than the signal? 
The answer to the first question is yes, the visibilities are within a few percent, same as the  EoR brightness temperature fluctuations for the cases considered.
Secondly, the in between region is not foreground free, due to non-negligible sidelobes. But, one can possibly extract signal from this region
if one models foregrounds and instruments accurately.
As all the three components, the cosmological signal, the foregrounds and the thermal noise are calculated in the same space, the analysis
suggested could be more straightforward.
\end{abstract}


\section{Introduction}
\label{sec:intro}
Epoch of Reionization is considered as one of the prime frontiers of observational cosmology. Evolution of the
universe during this epoch can be studied through the redshifted 21~cm line signal originating from the neutral hydrogen (HI) in the intergalactic medium (IGM) \citep[for reviews, see][]{2006ARA&A..44..415F,2006PhR...433..181F,2009CSci...97..841C,2012RPPh...75h6901P}. The observation of this cosmological signal is hindered due to the presence of much brighter astrophysical foregrounds \citep[see, e.g.,][]{2006PhR...433..181F,2010MNRAS.409.1647J} which are expected to dominate the HI signal by a factor of  $10^4\mbox{-}10^5$ \citep{2002ApJ...564..576D,2003MNRAS.346..871O,2004MNRAS.355.1053D,2008MNRAS.385.2166A}. 
The foregrounds are expected to be smooth functions of frequency while the signal, as it traces HI distribution in the universe, is expected
to oscillate rapidly with respect to the frequency. This principle has been used by experiment EDGES 
\footnote{https://www.haystack.mit.edu/ast/arrays/Edges/} \citep{2018Natur.555...67B}
 in their analysis for extracting the
global signal. To detect the fluctuating signal two different approaches has been proposed.
 One can take out the foreground signal by carefully modelling its frequency dependence \citep{2005ApJ...625..575S,2006ApJ...638...20B,2006ApJ...650..529W,2008MNRAS.391..383G,2009MNRAS.398..401L,2009MNRAS.394.1575L,
2009MNRAS.397.1138H,2010MNRAS.405.2492H,2011PhRvD..83j3006L,2011MNRAS.413.2103P,2012MNRAS.423.2518C,2015MNRAS.447.1973B,2015MNRAS.452.1587G}.   
This is approach has been called as Foreground Subtraction or Foreground Removal. 
It has been used to detect the 21cm signal by LOFAR (Low Frequency Array) \footnote{http://www.lofar.org} 
\citep{2017ApJ...838...65P,2018MNRAS.478.3640M,2019MNRAS.483.2207M,2019MNRAS.488.4271G}.
One can also measure the signal in the Fourier domain by focusing on
the region where signal is expected to be much large than the Foreground component \citep{2010ApJ...724..526D,2012ApJ...745..176V,2012ApJ...752..137M,2012ApJ...757..101T,2012ApJ...756..165P,2013ApJ...768L..36P,2013ApJ...770..156H,2014PhRvD..90b3018L,2014PhRvD..90b3019L,2015ApJ...804...14T}. The foregrounds, being smooth, tend to dominate in a wedge-shaped region in the $k_{\perp} \mbox{-} k_{\parallel}$ space, and hence the remaining portion of $k$-space ends up being nearly foreground free and is called as EoR window. Here $k_{\parallel}$ and $k_{\perp}$ are magnitudes of the Fourier modes in the directions parallel and perpendicular to the line of sight (LOS).
This method, called as foreground avoidance, is already being used by experiments like PAPER (Precision Array for Probing the Epoch of Reionization)\footnote{http://eor.berkeley.edu/} \citep{2015ApJ...809...61A} and MWA (The Murchison Widefield Array)\footnote{http://www.mwatelescope.org/} \citep{2016ApJ...833..213P,
2016ApJ...833..102B}. 
\par
\begin{figure}
\label{fig:sketch}
\includegraphics[width=0.4\textwidth]{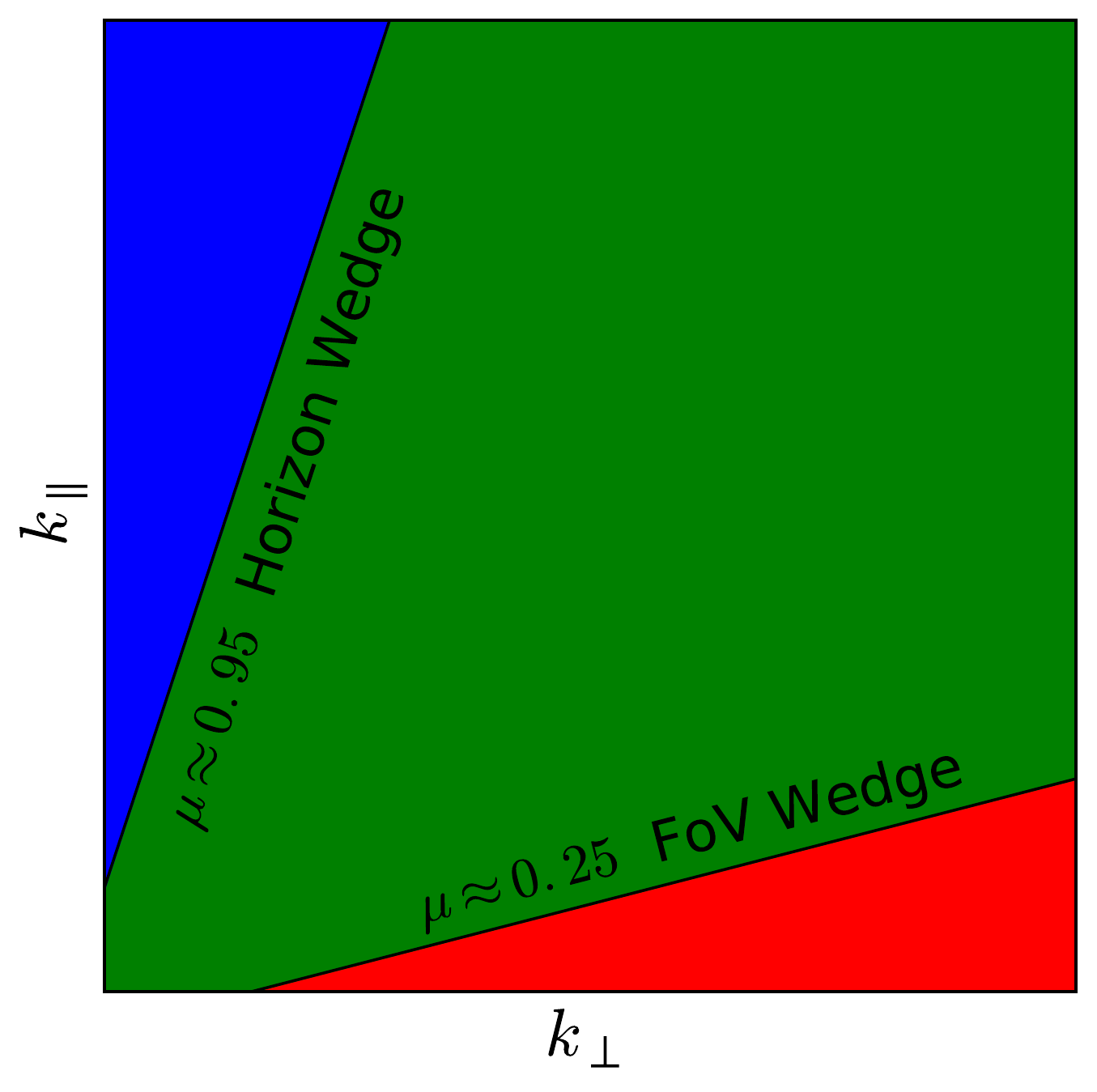}
\caption{Sketch indicating positions of FoV wedge and Horizon wedge for Telescope like SKA.}
\end{figure}

The extent of Foreground Wedge or EoR window is set by the geometric delay corresponding to the Horizon.
If one carefully considers the foregrounds in the 
$k_{\perp} \mbox{-} k_{\parallel}$ space, the situation is more like one depicted in the Figure~\ref{fig:sketch}.
There are three distinct regions, the wedge shaped region that lies below the line corresponding to the FoV wedge and which is dominated by foregrounds, the EoR window that lies above the line corresponding to the Horizon wedge and which is nearly Foreground free and the region in between
the two lines. 
Both the above mentioned telescopes, MWA and PAPER, have a very large Field of View (FoV) and do not have a very large number of baselines.
MWA and PAPER focus on shorter baselines as against SKA which has much longer baselines. So MWA and PAPER automatically
end up focusing on the blue region of the Figure~\ref{fig:sketch} as $k_{\perp}$ is proportional to the baseline distance.
Telescope like SKA has a smallish FoV (About 4 deg at $z=9.0$) which means that the FoV wedge occupies a smaller area in the 
$k_{\perp} \mbox{-} k_{\parallel}$ space. 
As $\Omega_{\rm FoV} \approx 4.8 \times 10^{-3} {\rm {sr}}$ is relatively smaller\footnote{$\Omega_{\rm FoV} = (\pi/4)(1.3\lambda/d_{\rm station})^2$}\footnote{http://astronomers.skatelescope.org/documents/{\small SKA-TEL-SKO-DD-001-1\_BaselineDesign1.pdf}} and as the number of baselines are large, the green region of the Figure~\ref{fig:sketch} that lies in between the two above mentioned lines is significant and can also yield good measurements of the signal provided the foregrounds are modelled properly and telescope calibration is done accurately.
\par
The core idea that is being presented in this paper is that instead of subtracting the foregrounds in the real space they can be modelled
and subtracted out in the Fourier space. The details of this procedure is given in the fifth section while the remaining sections give quantitative 
footing to this idea. It is expected that with careful modelling of foregrounds and telescope elements, one can achieve a good Signal to 
Noise ratio in the green region of the Figure\ref{fig:sketch} and utilize that region for measuring the signal.
The author is also of the opinion that it would be almost impossible to build images and do foreground subtraction in the real space unless the foreground parameters are
determined at accuracy of about 1 part in 10-100 thousand. This is because to build images one needs all the Fourier modes accurately measured. If the signal is
not correctly measured in some large and continuous portion of Fourier space (for example the Foreground Wedge region) then it would be impossible to reconstruct it by doing an inverse
Fourier transform. For a faithful construction of the signal one would have to measure it accurately all over the Fourier space. And to measure it accurately
in the Foreground wedge region, one would have to constrain foreground parameters to one part in 10-100 thousand or so as the signal is weaker by a factor of
10-100 thousand.
\par 
The paper is organized as follows: In section 2, I have discussed how to accurately get the visibilities corresponding to the EoR signal from the $k$-space brightness temperature 
fluctuations. In the third section, calculations of the second section are applied to a simulated cube obtained using 21cmFAST \citep{2011MNRAS.411..955M}. In the fourth section, an account of the calculations of the thermal noise in $k$-space is given. One is expected to compare this noise with the signal. In
the fifth section, the noise associated with Foreground modelling is considered. This can again compared to the
signal and thermal noise estimates. Throughout this paper, I have plotted various quantities in the $k_{\perp} \mbox{-} k_{\parallel}$ space. The cosmological parameters used for this study are $\Omega_m = 0.308$, $\Omega_{\Lambda} = 1 - \Omega_m$, $h = 0.678$, $\Omega_b = 0.049$, $\sigma_8 = 0.84$ and $n_s = 0.966$ \citep{2016A&A...594A..13P}.

\section{Visibilities to 21cm brightness temperature fluctuations} \label{sec:theory}
The fundamental equation of radio astronomy \citep{2017isra.book.....T,1999ASPC..180.....T} that connects the intensity distribution as a function of $\pmb{\theta}$
on sky to visibilities measured as a function baselines is given by,
\begin{equation}
V({\mathbf U},\nu) = \int_{\Omega} d^2\theta A_{\nu}({\pmb{\theta}}) I_{\nu}({\pmb {\theta}}) e^{- 2 \pi i {\mathbf U} \cdot {\pmb {\theta}}}
\label{eq:funda}
\end{equation}
Here, $\nu$ is frequency, $\Omega$ is Field of View (FoV), ${\mathbf U}$ is  baseline distance measured 
in units of wavelength $\lambda$ and $A_{\nu}({\pmb{\theta}})$ is the primary beam.
As far as EoR is concerned, this equation accomplishes Fourier Transforms with respect to two out of three co-ordinates. The two co-ordinates are
distances perpendicular to the LOS or the Line of Sight, (${\mathbf r}_{\perp}$) and the remaining third co-ordinate is the distance 
along LOS, $r_{\parallel}$. For small displacements along LOS, the frequency is linearly related to the LOS distances
and so a Fourier Transform of visibilities with respect to the frequency \citep{2004ApJ...615....7M} gives a representation of Intensity in the Fourier or $k$-space.
\begin{eqnarray}
V({\mathbf U},\eta) &=& \int_{B} d\nu V({\mathbf U},\nu) e^{- 2 \pi i \nu \eta} \\
&=& \int_{\Omega,B} d^2 \theta d\nu A_{\nu}({\pmb{\theta}}) I_{\nu}({\pmb {\theta}})  e^{- 2 \pi i \left( {\mathbf U} \cdot {\pmb{\theta}} + \nu \eta \right)} 
\label{eqn:deltra}
\end{eqnarray}
The quantity $|V({\mathbf U},\eta)|^2$ is directly related to the Power Spectrum of 21cm brightness temperature fluctuations with 
the components of wavevector $k_{\parallel}$ (parallel to LOS) and ${\mathbf k}_{\perp}$ (perpendicular to LOS) 
are directly proportional to the parameter $\eta$ and baselines ${\mathbf d}={\mathbf U} \times \lambda$, respectively . 
As shown in the appendix the calculations yields,
\begin{eqnarray*}
V({\mathbf k}_{\perp},k_{\parallel}) &=& \frac{1}{r'_{\parallel}D_c(z)^2} \int d r_{\parallel} d^2r_{\perp} I(r_{\parallel},{\mathbf r}_{\perp}) \exp 
\left[-i \left ( {\mathbf r}_{\perp} \cdot {\mathbf k}_{\perp} + r_{\parallel} k_{\parallel} \right) \right] \\
&\times& \exp \left( - 2 \pi i \frac{{\mathbf r}_{\perp} \cdot {\mathbf d}}{D_c(z) c} \frac{r_{\parallel}}{r'_{\parallel}} \right)
\end{eqnarray*}
\begin{eqnarray*}
{\mathbf k}_{\perp} &=& \frac{2 \pi}{D_c(z)} {\mathbf U} \\
k_{\parallel} &=& \frac{2 \pi}{r'_{\parallel}} \eta 
\end{eqnarray*}
\vspace{-0.25cm}
\begin{equation}
r'_{\parallel} = \frac{c(1+z)^2}{H(z)\nu_{21}} 
\label{eq:uetaconversiontok}
\end{equation}
Above equation assumes a flat beam and is used for calculating the 21cm power spectrum. This is fine as the aim is
to study the effect of the additional term, $\exp \left( - 2 \pi i \frac{{\mathbf r}_{\perp} \cdot {\mathbf d}}{D_c(z) c} 
\frac{r_{\parallel}}{r'_{\parallel}} \right)$, that is arising in the expression of power spectrum. 
For the 
case of computing power arising due to Foregrounds, I have assumed an Airy pattern beam (refer Figure~\ref{Airy}).    
\par
As seen from the exact calculations, the expression for $V({\mathbf U},\eta)$ contains a extra factor of phase 
which I would call as h-function.
\begin{equation}
h({\mathbf r}_{\perp},r_{\parallel}) \equiv \exp \left( - 2 \pi i \frac{{\mathbf r}_{\perp} \cdot {\mathbf d}}{D_c(z) c} \frac{r_{\parallel}}{r'_{\parallel}} \right)
\label{eq:hfunc}
\end{equation}
For short baselines ($|{\mathbf d}| \approx 0$) the phase is almost zero and so the h-function is unity. And hence
the Fourier transformed visibilities are simply Fourier transformed intensity fluctuations whose square is directly
proportional to the 21cm power spectrum. 
For a telescope like SKA1-Low the baselines range from a low of 35 metres, which is same as a station diameter, to a maximum of about 65km.
As we have an additional factor of h-function 
in the expression, we have,
\begin{equation}
V({\mathbf k}_{\perp},k_{\parallel}) \equiv {\tilde {\delta}}_{21}({\mathbf k}_{\perp},k_{\parallel}) \otimes {\tilde h}({\mathbf k}_{\perp},k_{\parallel})
\label{eq:visibs}
\end{equation}
That is Fourier transformed visibilities are proportional to the convolution of ${\tilde {\delta}}_{21}$, 21cm brightness temperature fluctuations in k-space 
and ${\tilde h}$, the Fourier transform of h-function for fixed $\mathbf{d}$. 
Note that theoretical 21cm power spectrum would correspond to $P_{21}({\mathbf k}_{\perp},k_{\parallel}) \equiv |{\tilde {\delta}}_{21}({\mathbf k}_{\perp},k_{\parallel})|^2$
while the one obtained in this approach would correspond to $P({\mathbf k}_{\perp},k_{\parallel})\equiv|V({\mathbf k}_{\perp},k_{\parallel})|^2$ \citep{2019MNRAS.483.2207M}.
From above equations one has,
\begin{equation}
P({\mathbf k}_{\perp},k_{\parallel}) \propto |{\tilde {\delta}}_{21}({\mathbf k}_{\perp},k_{\parallel}) \otimes {\tilde h}({\mathbf k}_{\perp},k_{\parallel})|^2
\end{equation}
\par
\begin{figure}
\includegraphics[width=0.4\textwidth]{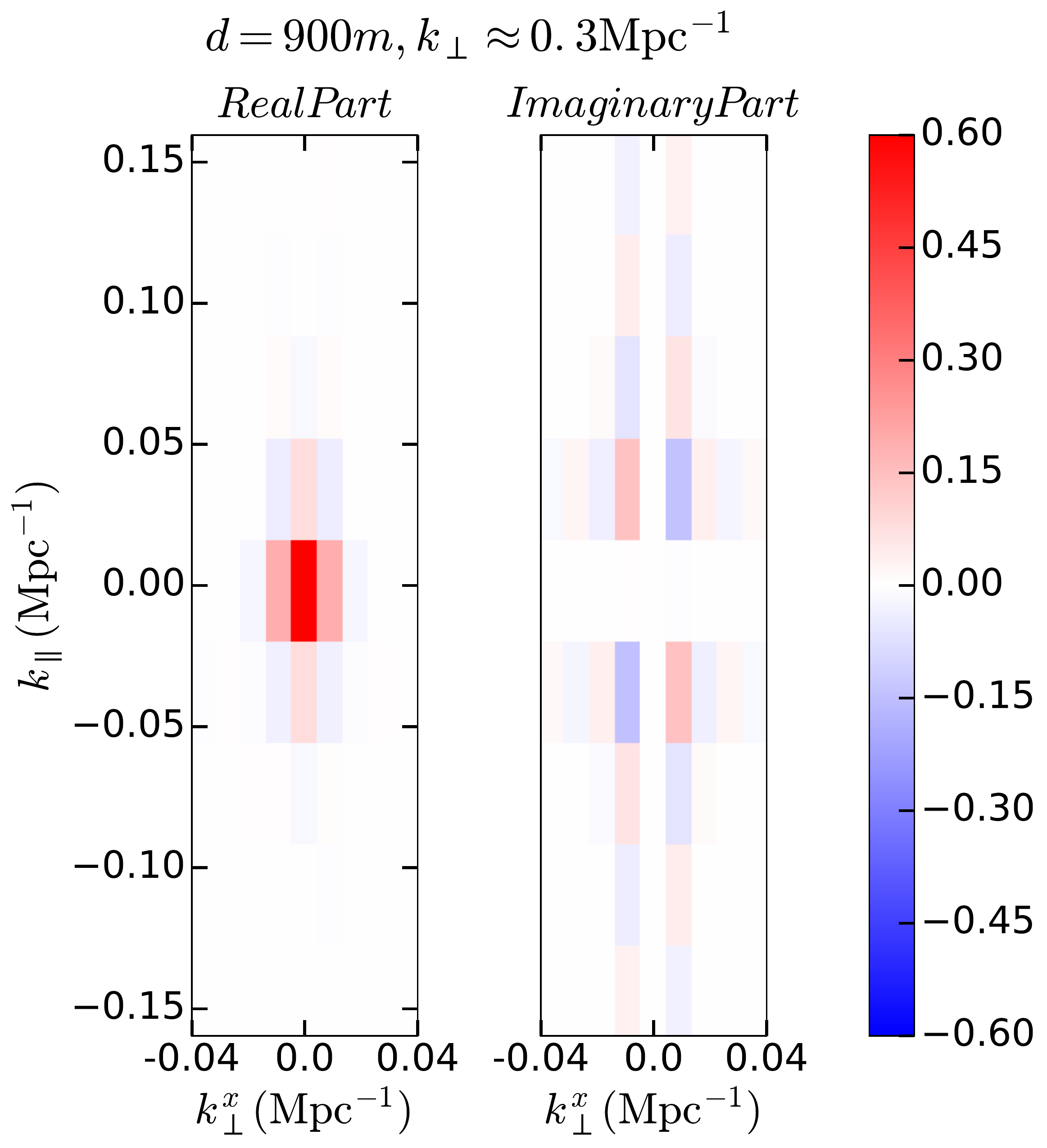}
\caption{Real and Imaginary parts of $\tilde{h}$-function for baseline of length $d=900m$. Horizontal axis is $k_x$, $x$-component
of $k_{\perp}$, while vertical axis is $k_{\parallel}$. The plot is for a baseline for which $d_x=d_y$.}
\label{fig:d900}
\end{figure}
\begin{figure}
\includegraphics[width=0.4\textwidth]{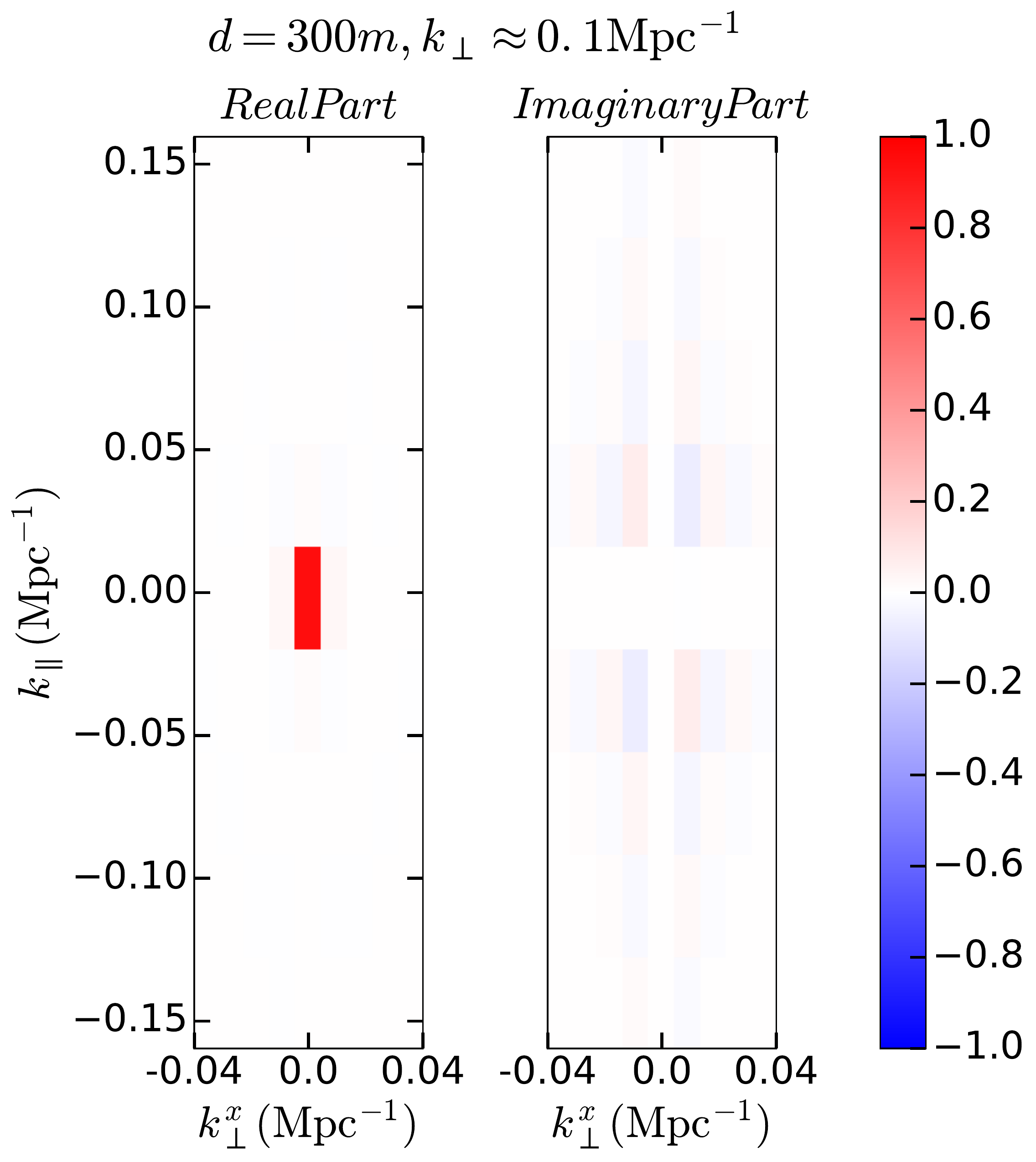}
\caption{Same as Figure~\ref{fig:d900} but for baseline of length $d=300m$.}
\label{fig:d300}
\end{figure}
\begin{figure}
\includegraphics[width=0.4\textwidth]{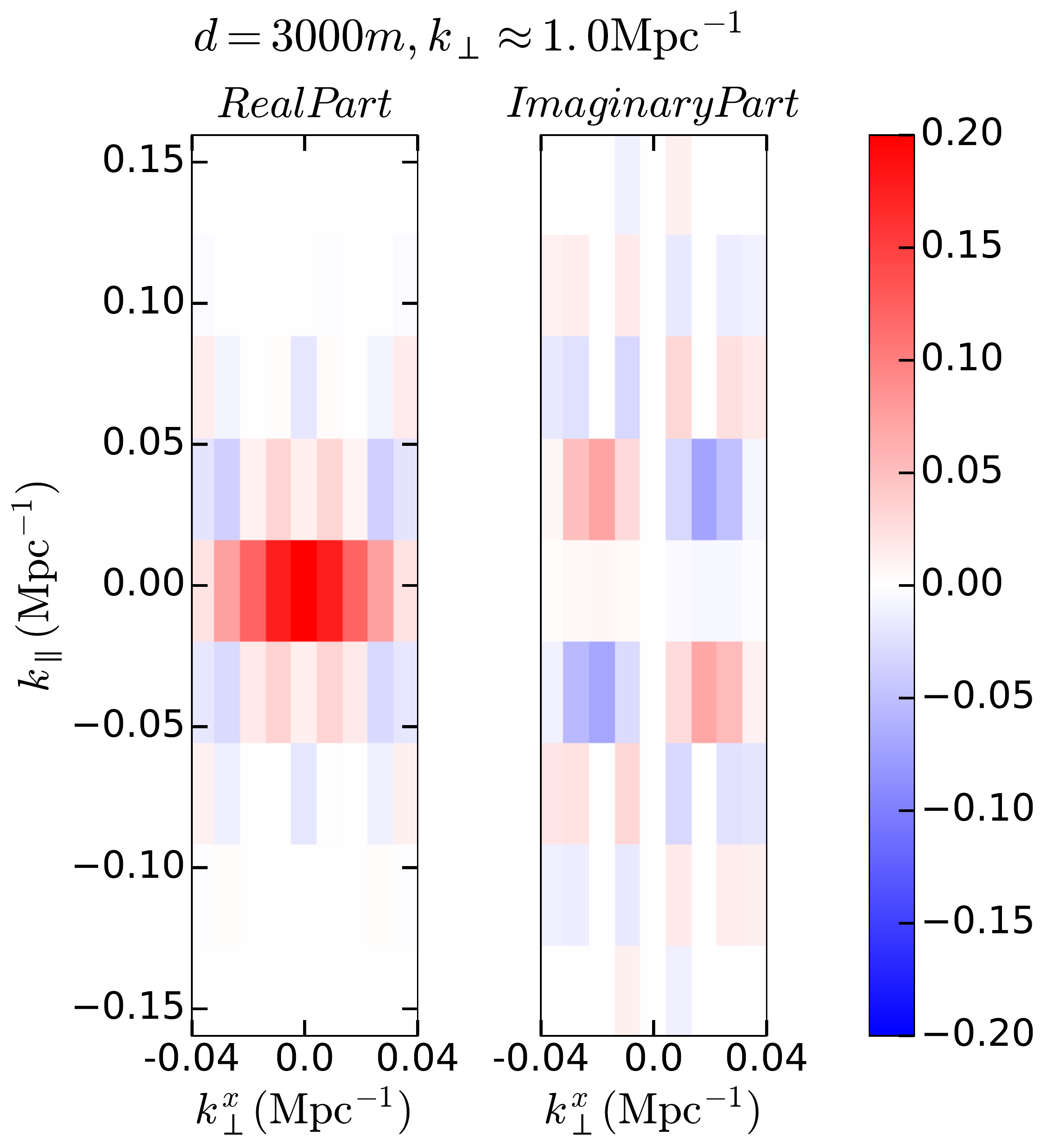}
\caption{Same as Figure~\ref{fig:d900} but for baseline of length $d=3000m$.}
\label{fig:d3000}
\end{figure}
\begin{figure}
\includegraphics[width=0.4\textwidth]{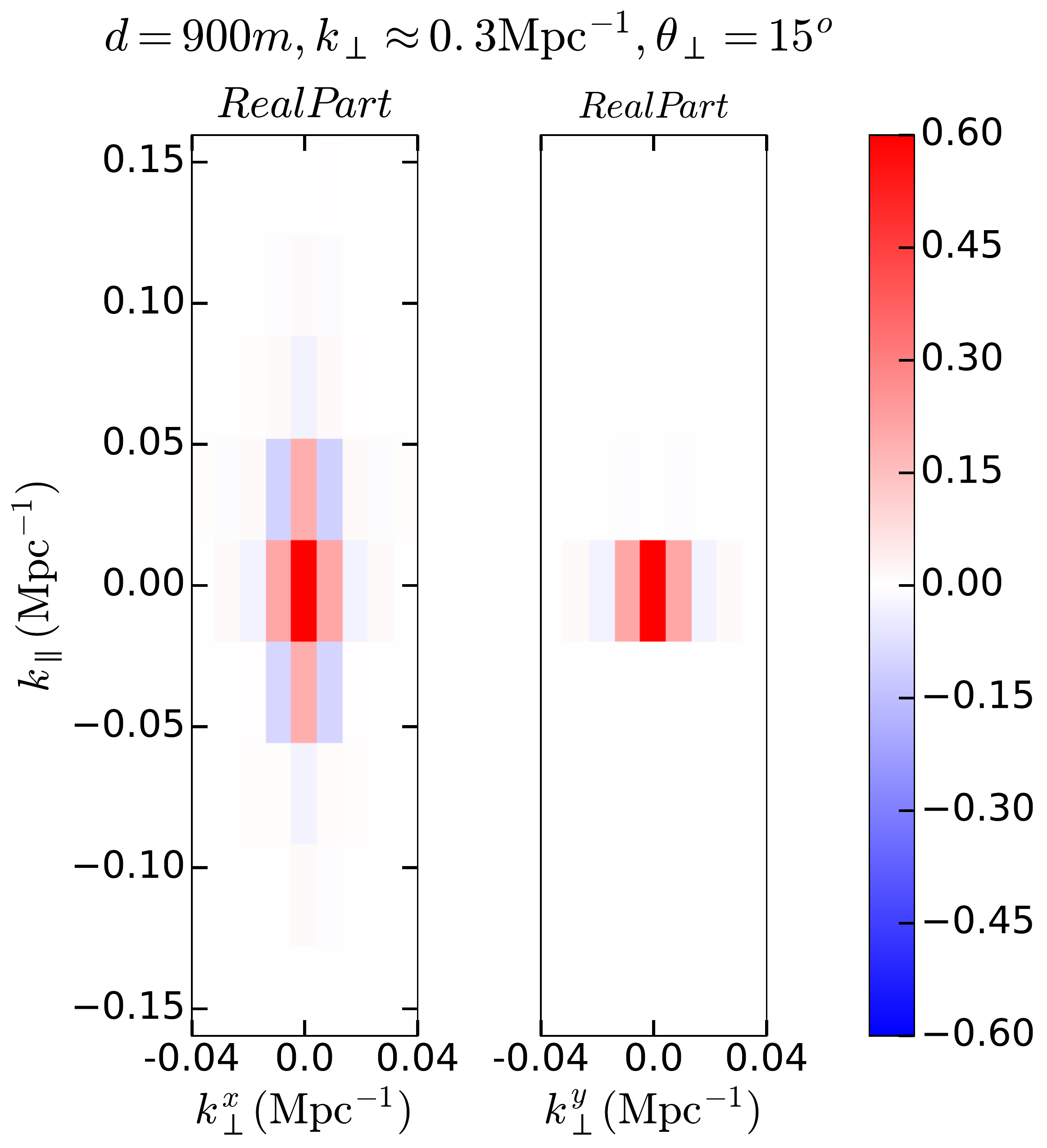}
\caption{Real part of $\tilde{h}$-function as function of $k_x$ and $k_y$ against $k_{\parallel}$.
The baseline $\mathbf{d}$ makes an angle of 15 degrees with respect to the $x$-axis.}
\label{fig:d900theta15}
\end{figure}
\begin{figure}
\includegraphics[width=0.4\textwidth]{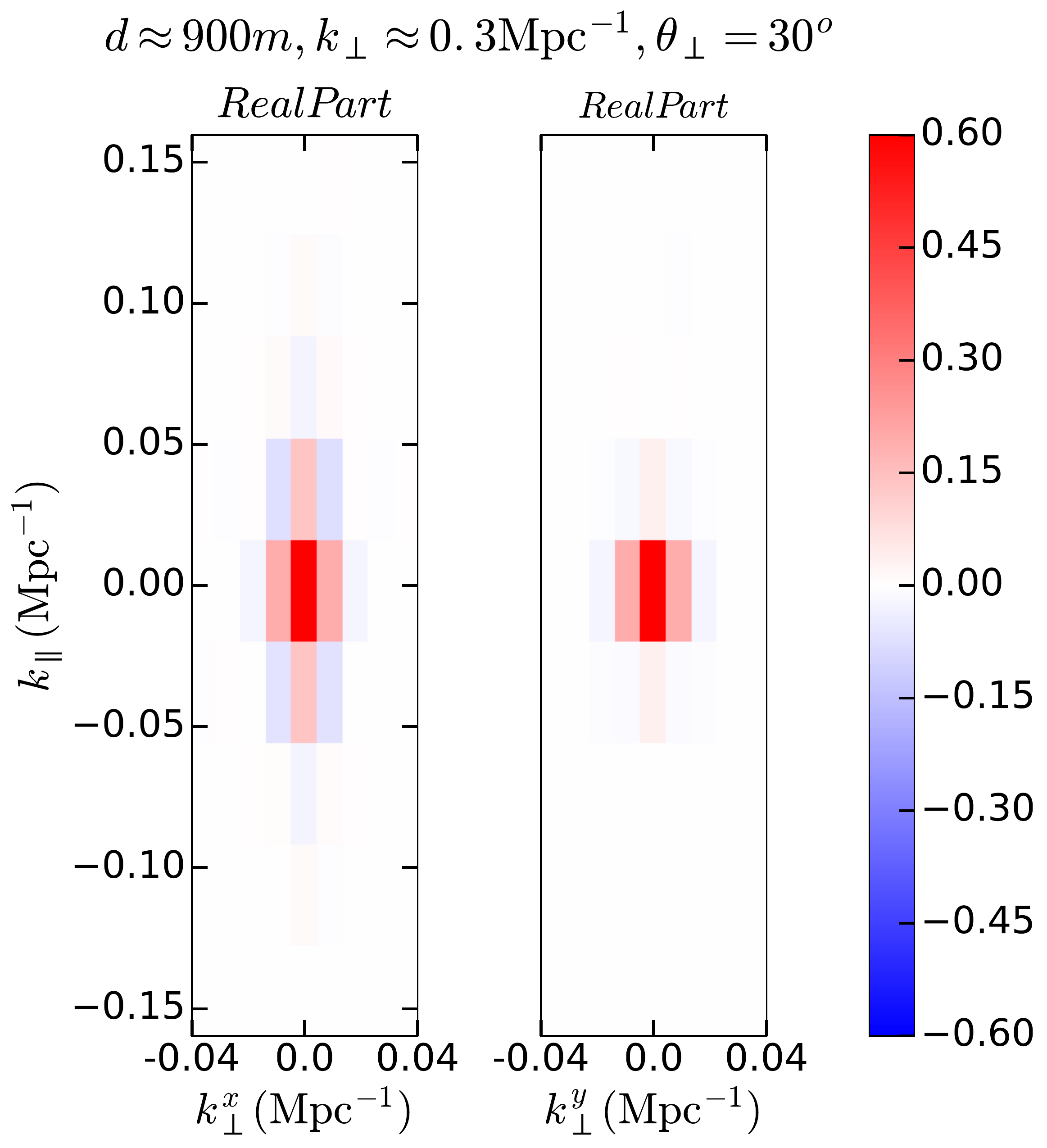}
\caption{Same as Figure~\ref{fig:d900theta15} but this time the angle is 30 degrees instead of 15.}
\label{fig:d900theta30}
\end{figure}

In this paper, it has been proposed that instead of trying to get the 21cm brightness temperature fluctuations by going to the real space, 
it could be worth exploring the option of doing the complete analysis in Fourier or ${\mathbf k}_{\perp}-k_{\parallel}$ plane.
 The Fourier transformed visibilities can be directly related to the 21cm brightness
  temperature fluctuations in the Fourier space and Foregrounds and other sources of noise can also be analyzed in the ${\mathbf k}_{\perp}-k_{\parallel}$ plane. 
As $h$-function is only a phase, summation of its Fourier transform gives zero imaginary part and unity for
real part. The ${\tilde h}$-function can be plotted for some baselines. Note that baseline also represent 
${\mathbf k}_{\perp}$ and for z = 9.0, that I consider in this paper, $d=300m$ roughly corresponds to 
${k}_{\perp} = 0.1 {\mathrm {Mpc}}^{-1}$. $k_x$ and $k_y$ are two components of ${\mathbf k}_{\perp}$. 
The real and imaginary parts of ${\tilde h}$-function for
3 representative baselines are plotted in Figures\ref{fig:d900}, \ref{fig:d300} and \ref{fig:d3000}.
The plots are for $d_x=d_y$. As expected, the real part is even under reflection 
while the imaginary part is odd under reflection. In the plots, $k_y =0$.
It is evident from Figure~\ref{fig:d300} that the ${\tilde h}$-function is almost like a delta function and 
so loss of power for the corresponding $k$-modes is small. As seen from Figure~\ref{fig:d3000}, the ${\tilde h}$-function is significant for non-zero arguments and this would cause a depreciation in the measured fluctuations and hence the power spectrum. 
The baselines components $x$ and $y$ can have different relationship in the baseline plane.
If $d_x$ and $d_y$ are located in such a way that they subtend an angle of 15 degrees with 
respect to the positive $x$-axis then one gets Figure~\ref{fig:d900theta15}
For 30 degrees angle one gets Figure~\ref{fig:d900theta30}.
Once we have obtained desired $\tilde{h}$-functions, the next step is to generate signal with the help of simulations
and see how the convolution affects the power measured in terms of visibilities.
The box-size used in this paper is 630Mpc as it corresponds to the approximate 4 degree FoV. The bandwidth used if 8MHz ($\approx 160\rm{Mpc}$) and 
channel width is about 62.5kHz (128 channels in the bandwidth). Baseline range is from about 100m to 3km which also corresponds to 
$k_{\perp}$ of about $.03{\mathrm{Mpc}}^{-1}$ to $1.0{\mathrm{Mpc}}^{-1}$.
\section{Signal Power Spectrum}
\label{sec:21cmfast}
To generate 21~cm brightness temperature cubes, I used publicly available package 21cmFAST 
\footnote{https://github.com/andreimesinger/21cmFAST/} \citep{2011MNRAS.411..955M}.
I have used box of size (630Mpc)$^3$
with grid-size of 256$^3$. The cosmological parameters were as mentioned before while the reionization parameters were $\zeta = 30$, $M_{\mathrm {min}} = 4.9 \times 10^8 \Msun$ and $R_{\mathrm{HII,max}} = 50{\mathrm {Mpc}}$. The redshift of computation was $z=9.0$. If one considers
a slice along the frequency direction of width about 8MHz, it would correspond to LOS distance of about 160Mpc. As our observation
slice is about one fourth of the observation width, one can average every four successive measurements along $k_{\parallel}$
and assign them to a mean $k_{\parallel}$-bin. This also gives $\Delta k_{\parallel} \approx 4 \Delta k_{\perp}$, as required.
The dimensionless 21~cm power spectrum is given as usual,
\begin{equation}
\left<{\tilde {\delta}}_{21}(\mathbf{k}) ~{\tilde{\delta}}_{21}^*(\mathbf{k'}) \right> = (2 \pi)^3~\delta_D(\mathbf{k} - \mathbf{k'})~P(\mathbf{k}),
\end{equation}
\begin{equation}
\Delta^2(\mathbf{k}) = \f{k^3 P(\mathbf{k})}{2 \pi^2}.
\end{equation}
The spherically averaged power spectrum is obtained by averaging $\Delta^2(\mathbf{k})$ over all possible angles
\begin{equation}
\Delta^2_0(k) = \f{1}{2} \int_{-1}^1 \der \mu~\Delta^2(\mathbf{k}) = \int_0^1 \der \mu~\Delta^2(\mathbf{k}),
\end{equation}
where $\mu=k_{\parallel}/k$ is the cosine of the angle that wavevector ${{\mathbf k}}$ makes with the LOS direction. The other $k$-component perpendicular to the LOS direction is denoted as $k_{\perp}$. The line that separates the Foreground dominated region from the remaining portion is represented by,
\begin{equation}
k_{\parallel} \leq C_{\mathrm{FoV}}~k_{\perp},~~
C_{\rm {FoV}} = \sin \theta_{\rm FoV}~\f{D_c(z) H(z)}{c (1+z)},
\label{eq:C}
\end{equation}
where $D_c(z)$ is comoving distance to redshift $z$, $H(z)$ is the Hubble constant at redshift $z$ and $\theta_{\rm {FoV}}$ the field of view in radians. In terms of $\mu$, the equation of line would be 
\begin{equation}
\mu^{\rm FoV}_{\rm min} = \f{C_{\rm FoV}}{\sqrt{1 + C_{\rm FoV}^2}}.
\label{eq:mumin}
\end{equation}
The EoR window is represented by the region $1 \ge \mu \ge \mu^{\rm{Hor}}_{\rm{min}}$ where
\begin{equation}
k_{\parallel} \leq C_{\mathrm{Hor}}~k_{\perp},~~
C_{\rm {Hor}} = \sin \theta_{\rm Hor}~\f{D_c(z) H(z)}{c (1+z)},
\label{eq:Hor}
\end{equation}
with $\theta_{\rm Hor} = 90^{o}$.
\begin{equation}
\mu^{\rm {Hor}}_{\rm min} = \f{C_{\rm {Hor}}}{\sqrt{1 + C_{\rm Hor}^2}}.
\label{eq:mumin}
\end{equation}
\par
\begin{figure}
\includegraphics[width=0.475\textwidth]{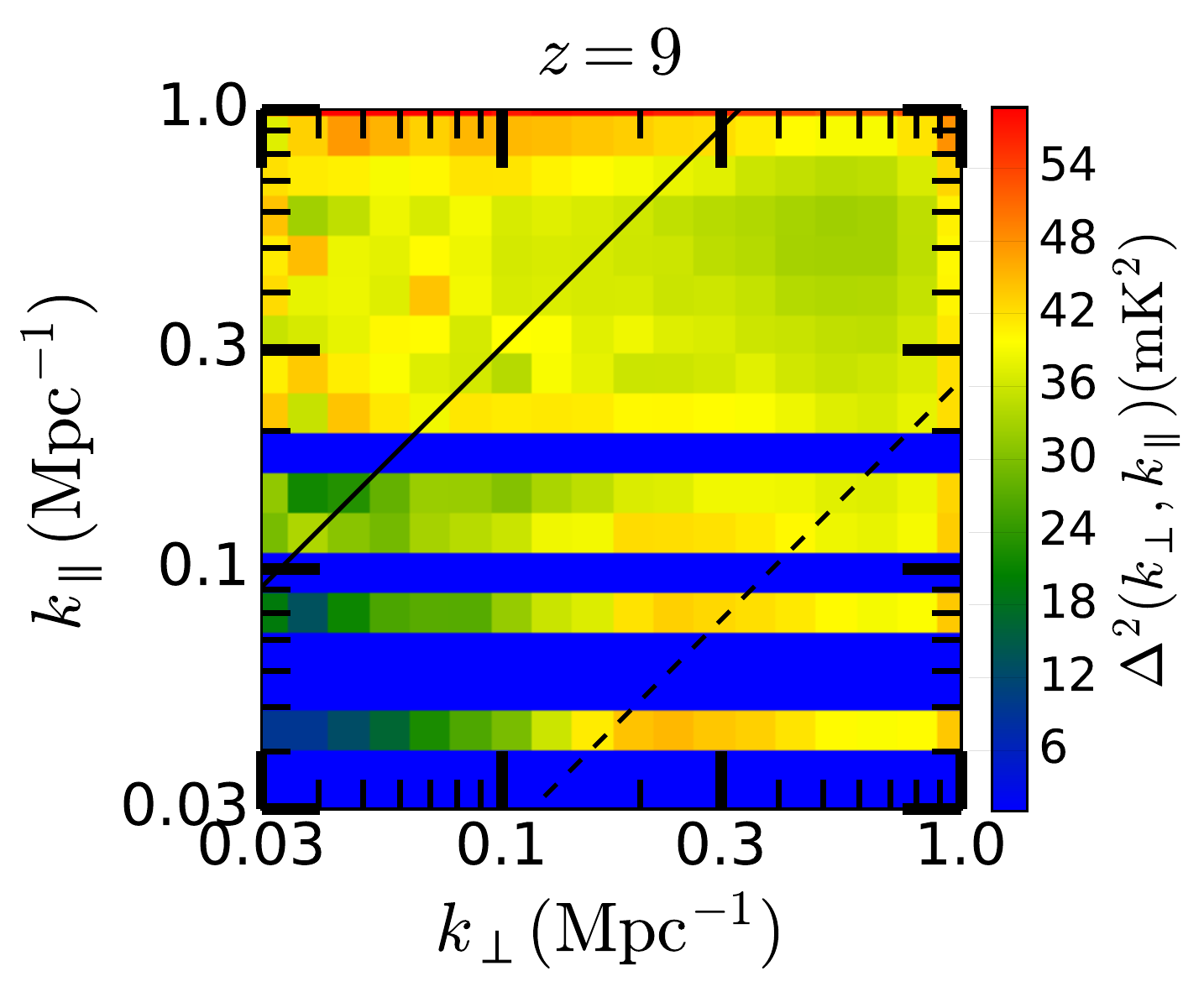}
\caption{The 21~cm signal power spectrum in the cylindrical $k_{\perp}\mbox{-}k_{\parallel}$ space. Horizontal blue stripes indicate
the region of $k$-space where measurements are unavailable. This is because LOS extent is about 160Mpc, about one fourth of the perpendicular extent which is about 630Mpc. The solid line corresponds to $\mu_{\rm {min}}=0.95$(approximately the Horizon wedge) while the dashed line corresponds to $\mu_{\rm{min}}=0.25$ (approximately the Field of View wedge).}
\label{fig:21cmpownohtilde}
\end{figure}
\begin{figure}
\includegraphics[width=0.475\textwidth]{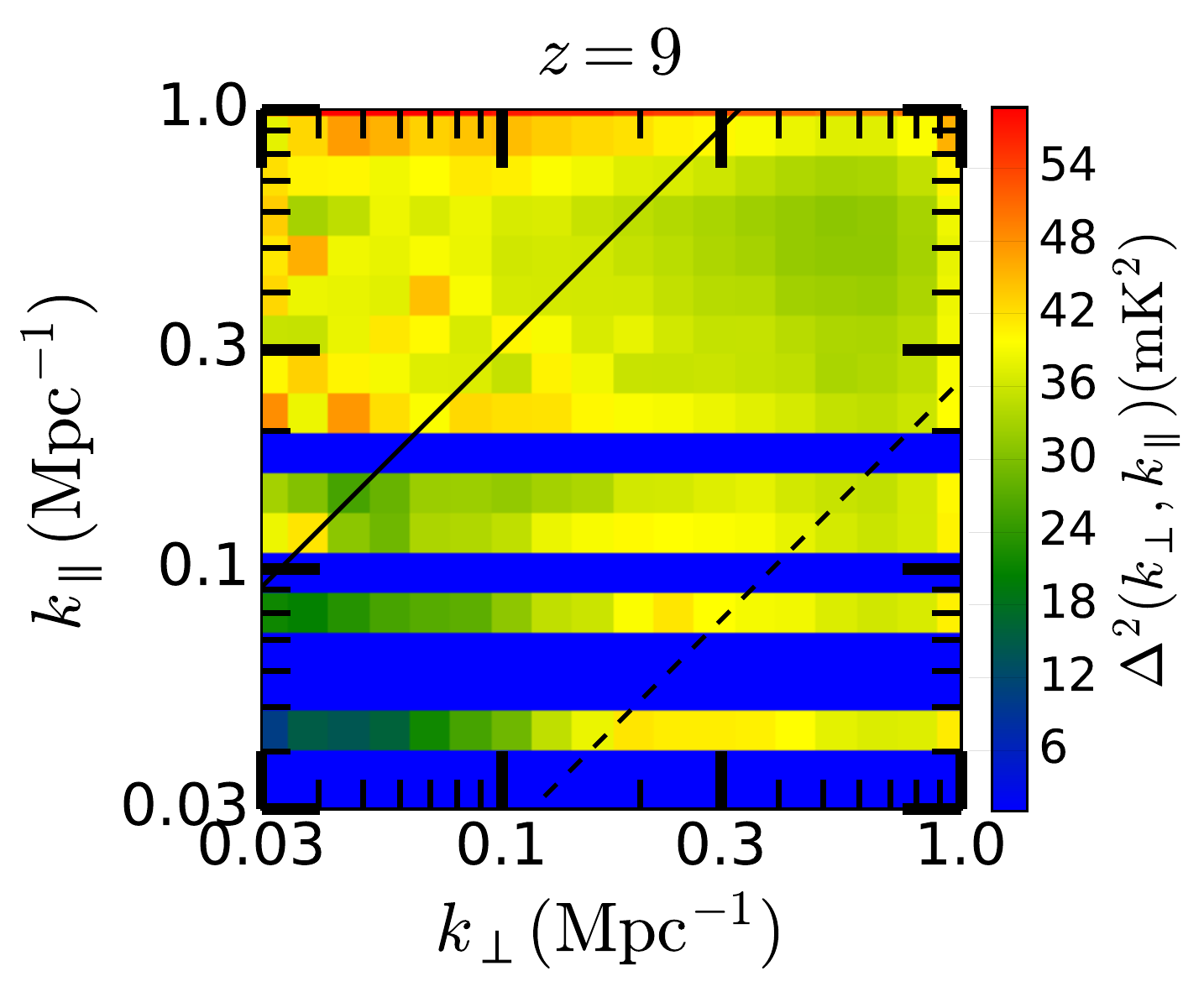}
\caption{The 21~cm signal power spectrum, this time obtained by convolving with the $\tilde{h}$-function. }
\label{fig:21cmpowwithhtilde}
\end{figure}
\begin{figure}
\includegraphics[width=0.475\textwidth]{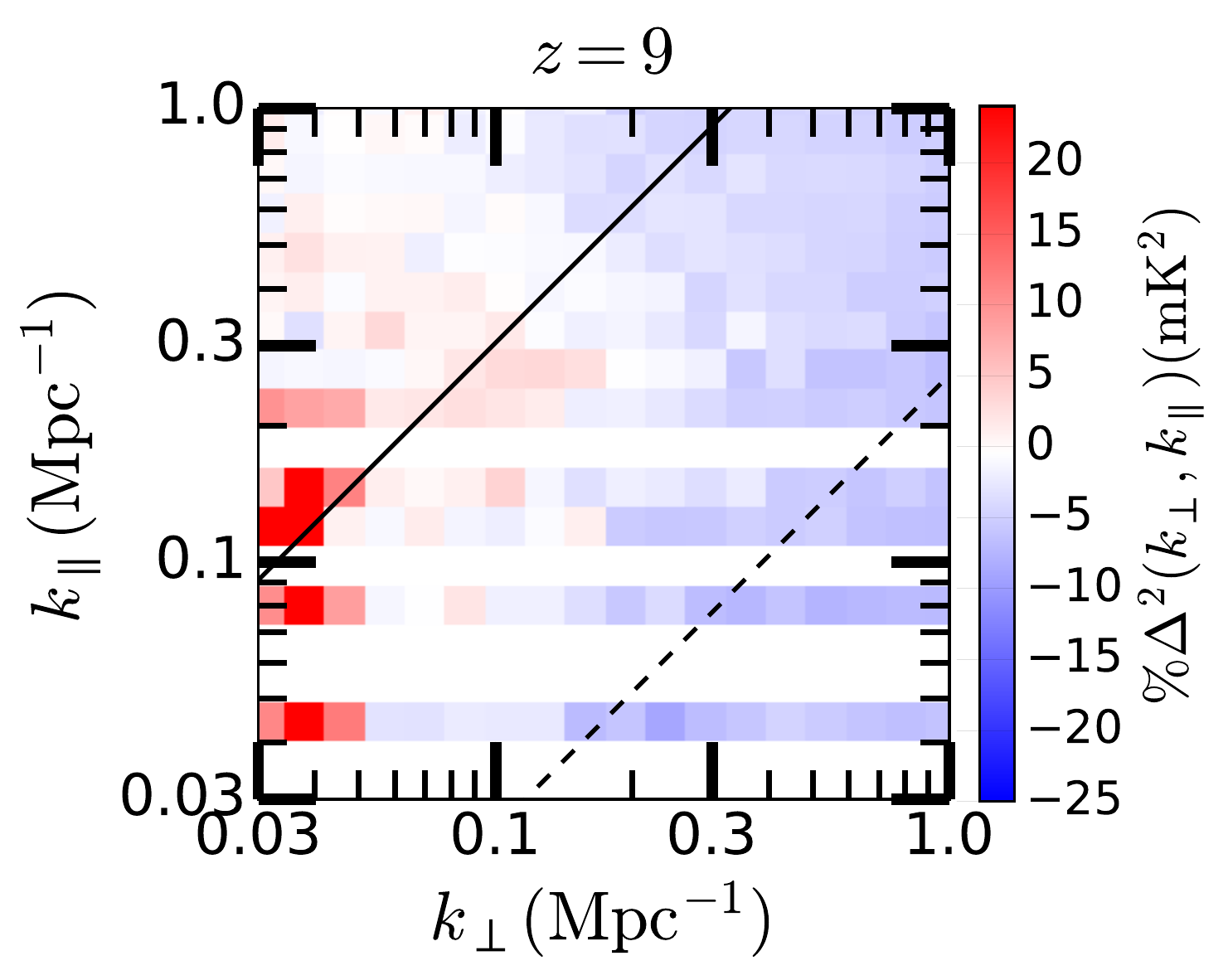}
\caption{The difference between true power (Figure~\ref{fig:21cmpownohtilde}) and power obtained after convolution (Figure~\ref{fig:21cmpowwithhtilde})}
\label{fig:diff}
\end{figure}

As mentioned earlier, we would like to extract as much information as possible from the region in between:  
$\mu^{\rm{Hor}}_{\rm{min}} \ge \mu \ge \mu^{\rm{FoV}}_{\rm{min}}$
After obtaining the equations for $\mu_{\rm {min}}^{\mathrm{FoV}}$ and $\mu_{\rm {min}}^{\rm{Hor}}$, I discuss how to get visibilities
corresponding to the 21cm brightness temperature distribution. 
As mentioned earlier, they are convolution of 21~cm fluctuations in the $k$-space with the $\tilde{h}$-function. To see the effect of
convolution, it is advisable to look at 21~cm fluctuations without any convolution. This is shown in Figure~\ref{fig:21cmpownohtilde}.
To get the convolved power, $d_x$ and $d_y$ were divided into 20 logarithmically spaced $d$-bins along both $x$ and $y$ directions.
Note that the $h$-function explicitly depends on the baseline ${\mathbf d}$.
The $h$-function and its Fourier transforms were computed for all these $20 \times 20 = 400$ bins separately. The $\tilde{h}$-function
varies only little within each of these bins and so was used to perform the convolution with the 21~cm fluctuations obtained from
21cmFAST. The convolved cube was used to get convolved power and it plotted in the Figure~\ref{fig:21cmpowwithhtilde}. 
The difference between the true power and the power obtained after convolution is plotted in Figure~\ref{fig:diff}.
It can be seen that there is up-to 5-10\% decrement in the power, especially for larger baselines.
A similar conclusion for the case of PAPER telescope was obtained earlier by \citet{2012ApJ...756..165P}, although the
approach used was different.
As one can see, the error rises for larger $k_{\perp}$. The errors could go to much larger 
values if one is looking at much larger $k_{\perp}$ ($\sim 10 {\mathrm{Mpc}}^-1$) or at similar $k_{\perp}$ ($0.1 - 1.0 {\mathrm{Mpc}}^{-1}$) 
for a telescope with much larger FoV.
There is somewhat excess power around $k_{\perp}=0.03{\mathrm{Mpc}}^{-1}$ and small $k_{\parallel}$. This is because the power
spectrum (or $\delta_{21}(k)$) is rising very sharply around $k \sim 0$ and convolution with respect to $\tilde{h}$-function is transferring power 
from small k-modes to larger k-modes.
After seeing the estimate of the signal, the next step is to look for thermal noise computation.  
\section{Thermal Noise}
\label{sec:thermal}
\begin{figure}
\includegraphics[width=0.5\textwidth]{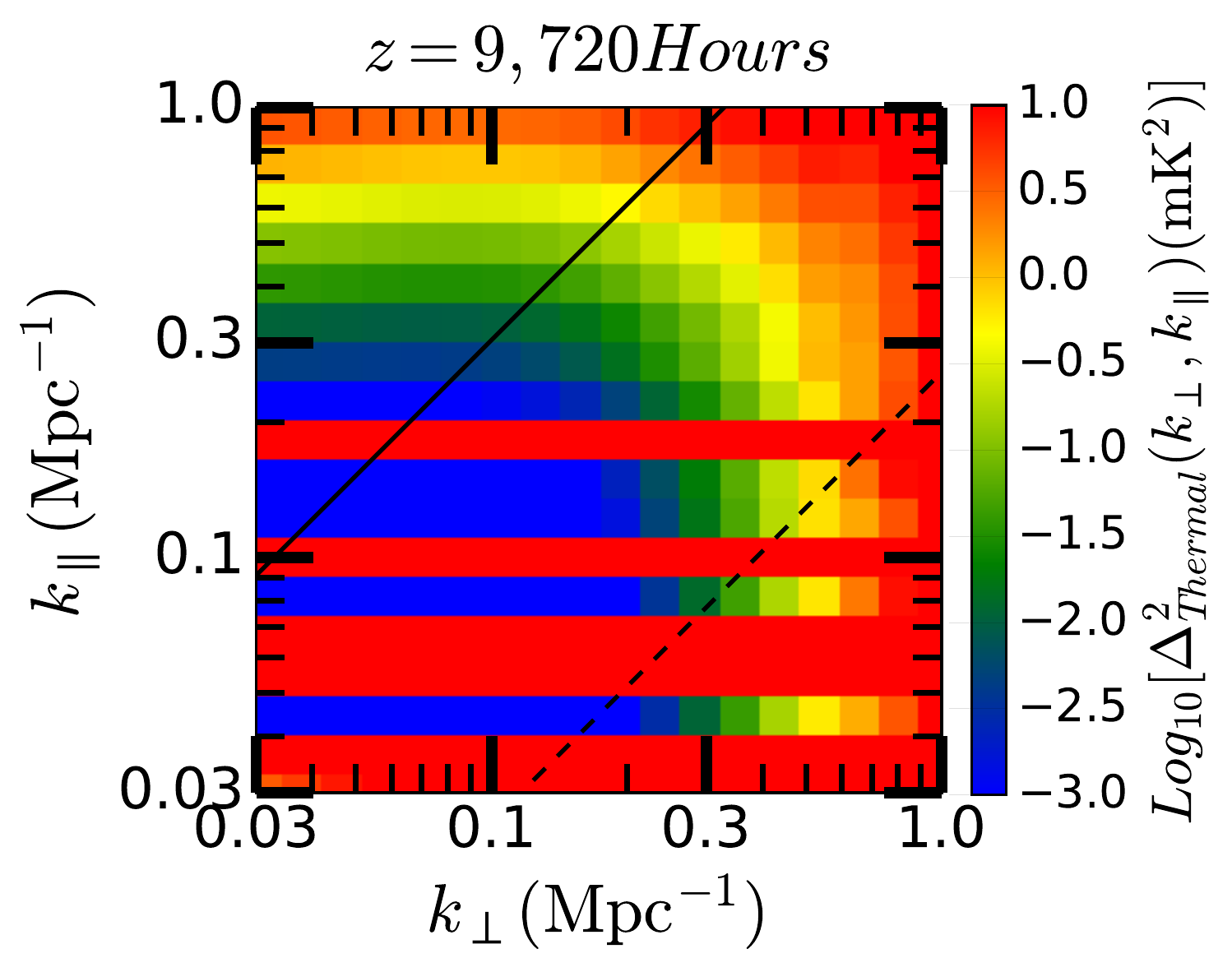}
\caption{Thermal noise in $\mathbf{k}_{\perp}-k_{\parallel}$ space for observing time of 720 hours (6 hours for 120 days).}
\label{fig:thermalnoise}
\end{figure}

\begin{figure}
\includegraphics[width=0.5\textwidth]{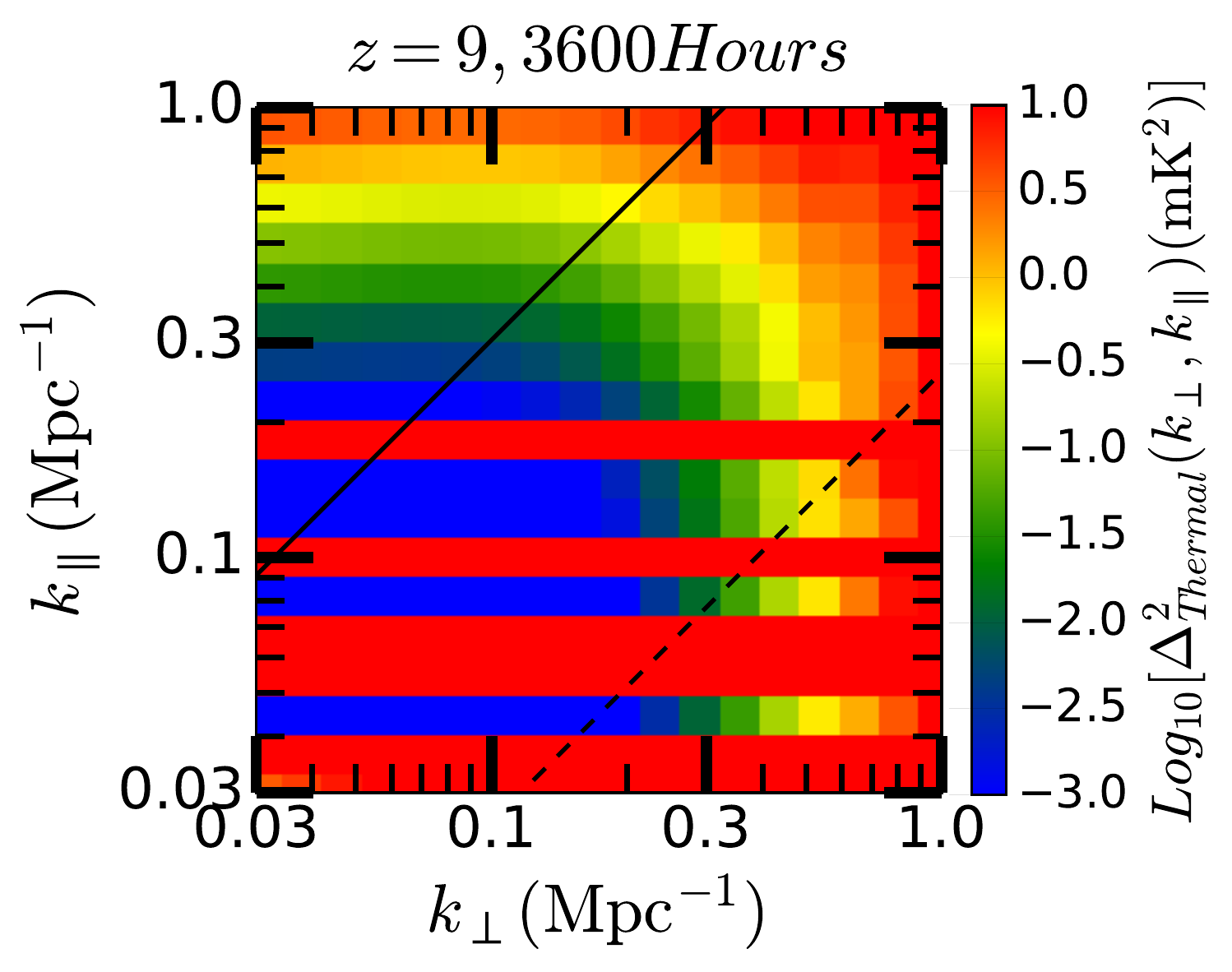}
\caption{Thermal noise in $\mathbf{k}_{\perp}-k_{\parallel}$ space for observing time of 3600 hours (6 hours for 600 days).}
\label{fig:thermalnoise5}
\end{figure}

System temperature needed for estimation of power due to thermal noise is given by
\begin{equation}
  T_\mathrm{sys}=60~\mathrm{K}\left(\frac{300~\mathrm{MHz}}{\nu_c}\right)^{2.55} + T_{rcvr},
\end{equation}
where $\nu_c=1420MHz/(1+z)$ is the observation frequency and $T_{rcvr} \approx 60 {\mathrm K}$\footnote{http://astronomers.skatelescope.org/documents/{\small SKA-TEL-SKO-DD-001-1\_BaselineDesign1.pdf}}
As described in \citet{2006ApJ...653..815M, 2012ApJ...753...81P} the thermal noise for each measured Fourier mode
is given by,
\begin{equation}
  \Delta_{\mathrm{thermal}}^2(k)\approx X^2Y\frac{k^3}{2\pi^2}\frac{\Omega}{2t}T^2_\mathrm{sys},
  \label{eqn:noisepermode}
\end{equation}
where $k=(k_{\perp}^2+k_{\parallel}^2)^{1/2}$ is absolute magnitude of the Fourier mode, $\Omega$ is the field of view of each interferometric element, $t$ is the integration time in seconds for the mode and $X$, $Y$ are cosmology dependent factors which are given as
\begin{equation}
  X\approx 1.9 \frac{h^{-1}\mathrm{cMpc}}{\mathrm{arcmin}} \left(\frac{1+z}{10}\right)^{0.2},
\end{equation}
and
\begin{equation}
  Y\approx 11.5 \frac{h^{-1}\mathrm{cMpc}}{\mathrm{MHz}} \left(\frac{1+z}{10}\right)^{0.5}\left(\frac{\Omega_mh^2}{0.15}\right)^{-0.5}.
\end{equation}
Equation~\ref{eqn:noisepermode} is for a single $k$-mode sampled continuously for time $t$. 
As the Earth rotates, baselines rotate and the values of $\mathbf{k}_{\perp}$ corresponding to the baselines also change as dictated by equations~\ref{eq:uetaconversiontok}. The extent of time for which any mode can be sampled continuously is given by \citep{2012ApJ...753...81P}.
\begin{equation}
t_{\rm {per\mbox{-}mode}} = t_{20} \left[ \frac{\Omega_0}{\Omega} \right]^{\frac{1}{2}} \left[ \frac{20}{|{\mathbf u}|} \right].
\label{eq:tpermode}
\end{equation}
Here $\Omega_0$ is PAPER Field of View while $\Omega$ is the FoV of the concerned telescope, in this case SKA. $t_{20}$ is time of continuous sampling
corresponding to a baseline of length $20 \times \lambda$. 
To get thermal noise for any mode for practical scenario one has to consider multiple factors. First one is that there would be
binning involved in $k$-space and each $k$-bin can contain multiple samples, say $N_{k{\mbox{-}}bin}$. Each mode in a given bin would also be
sampled multiple times in a day. This quantity is $N_{t}=t_{per{\mbox{-}}day}/t_{per{\mbox{-}}mode}$. Total samples for any $k$-bin
would be $N_{k{\mbox{-}}bin} \times N_{t}$. Considering all these factors, 
one can arrive at following expression for thermal noise for each $k$-bin
\begin{equation}
\Delta_{\mathrm{thermal}}^2({\mathbf{k}})\approx X^2Y\frac{k^3}{2\pi^2}\frac{\Omega}{2~t_{\rm{per\mbox{-}mode}}~t_{\rm{days}}} {T^2_\mathrm{sys}} \times \f{1}{(N_{\rm{per\mbox{-}mode}})^{1/2}},
\label{eq:thermalnoise}
\end{equation}
where $N_{\rm{per\mbox{-}mode}}=N_{k{\mbox{-}}bin} \times N_{t}$ and $t_{days}$ is the number of days telescope is observing. Note that
a square root appears on $N_{\rm{per\mbox{-}mode}}$ as here the modes are added incoherently while there is no square root on 
$t_{\rm{per\mbox{-}mode}}$ as the mode is sampled coherently for this duration.
Figure~\ref{fig:thermalnoise} plots this thermal noise in the cylindrical $k$-space for observing time of 6 hours each for 120 days.
While, the Figure~\ref{fig:thermalnoise5} plots the same quantity for a time period 5 times as big.
\section{Foreground Noise}
\label{sec:FG}
As mentioned in earlier sections, the main reason the EoR signal could be extracted is the fact that it 
oscillates rapidly as a function of frequency while the Foreground component is smooth as function of frequency.
So to get the Foreground component, the idea is to smooth the total signal over fairly wide frequency interval ($\ge 5MHz$)
 so that the 
fluctuating signal and the mode-mixing components would average 
out to a very small value. Then fit to the smoothed signal a physically motivated foreground model. 
In the case of statistical signal, the foreground fitting has to be
done along each and every Line of Sight. That is, when one obtains image cube from smoothed visibilities, the telescope
resolution would naturally yield a certain possible number of Line of sights through this cube. And the foreground
model has to be fitted to each Line of Sight separately assuming that every LOS traces slightly different section of the sky
\citep{2010MNRAS.409.1647J}. 
Such a procedure would yield foreground component but with its parameters along each LOS having some uncertainties.
Once we have a foreground model, it can be used to simulate visibilities and power corresponding to this model 
(Equation~\ref{eq:funda}). Uncertainties in the parameters of the foreground model would give rise to Foreground Noise
in the measurement of the signal power spectrum. 
There would be additional sources of errors due to  
instrumental mode mixing \citep{2016ApJ...820...51P, 2018MNRAS.478.3640M, 2018MNRAS.474.4552G}.  The instrumental mode mixing 
would generate additional noise and that can be added in quadrature to the other two components of noise, thermal and 
foregrounds.
Simulations of power generated in the Fourier space due to mode-mixing would require simulating the visibilities of the mode-mixing 
processes. 
A thorough treatment 
of the errors and power spectra of mode-mixing components on the lines presented in the above mentioned papers is beyond the scope of this paper. 
But, it could be taken up as future work.
In this paper, this Foreground noise is simulated for some particular cases. We defer the detailed analysis for 
all the other cases for future work. This paper shows that this FG noise component is small enough to give a reasonable signal to noise ratio.
\par
\begin{figure}
\includegraphics[width=0.5\textwidth]{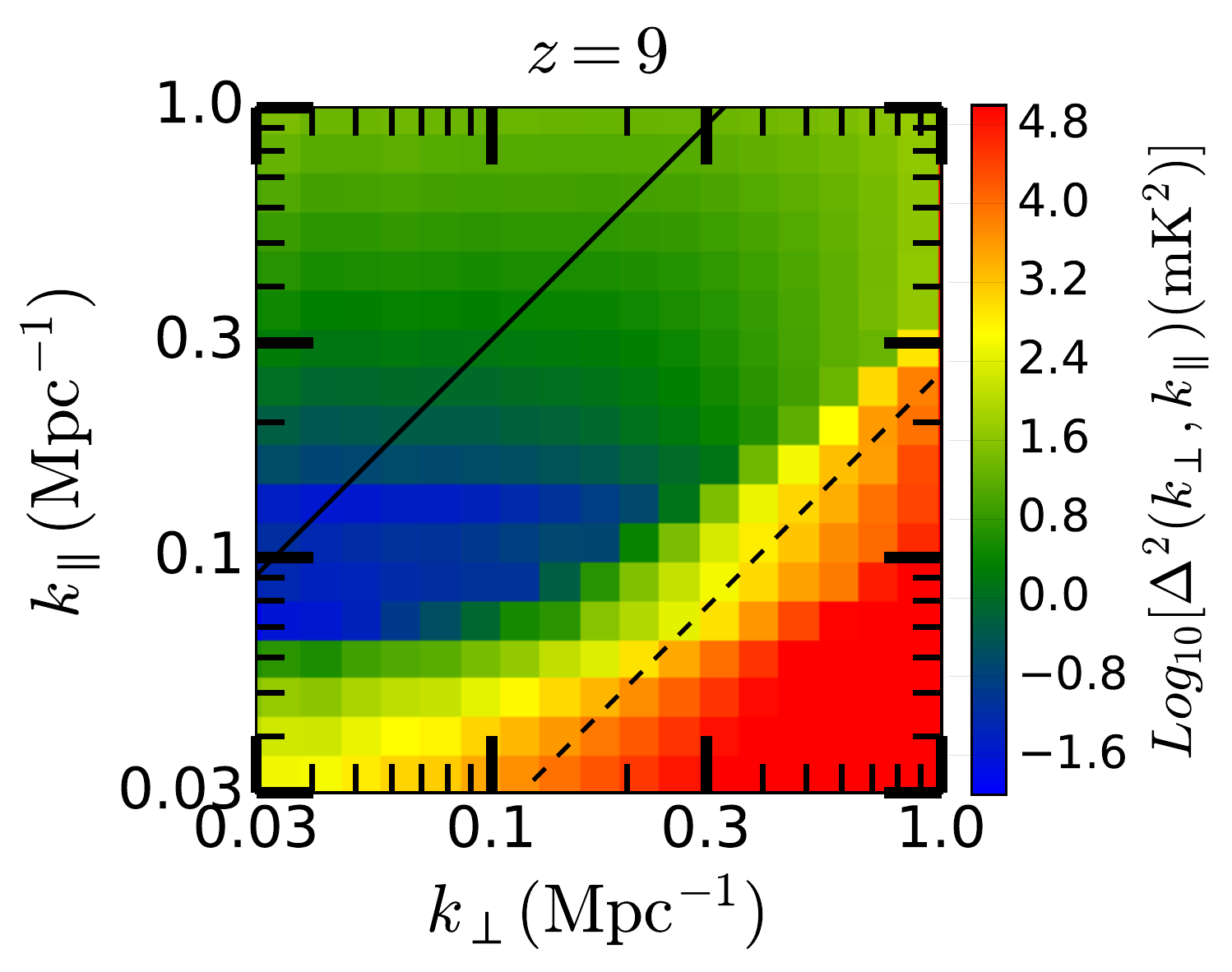}
\caption{Foreground noise in $k_{\perp}\mbox{-}k_{\parallel}$ space for 2\%uncertainty in the amplitude of GDSE.}
\label{FGAmp}
\end{figure}
\begin{figure}
\includegraphics[width=0.5\textwidth]{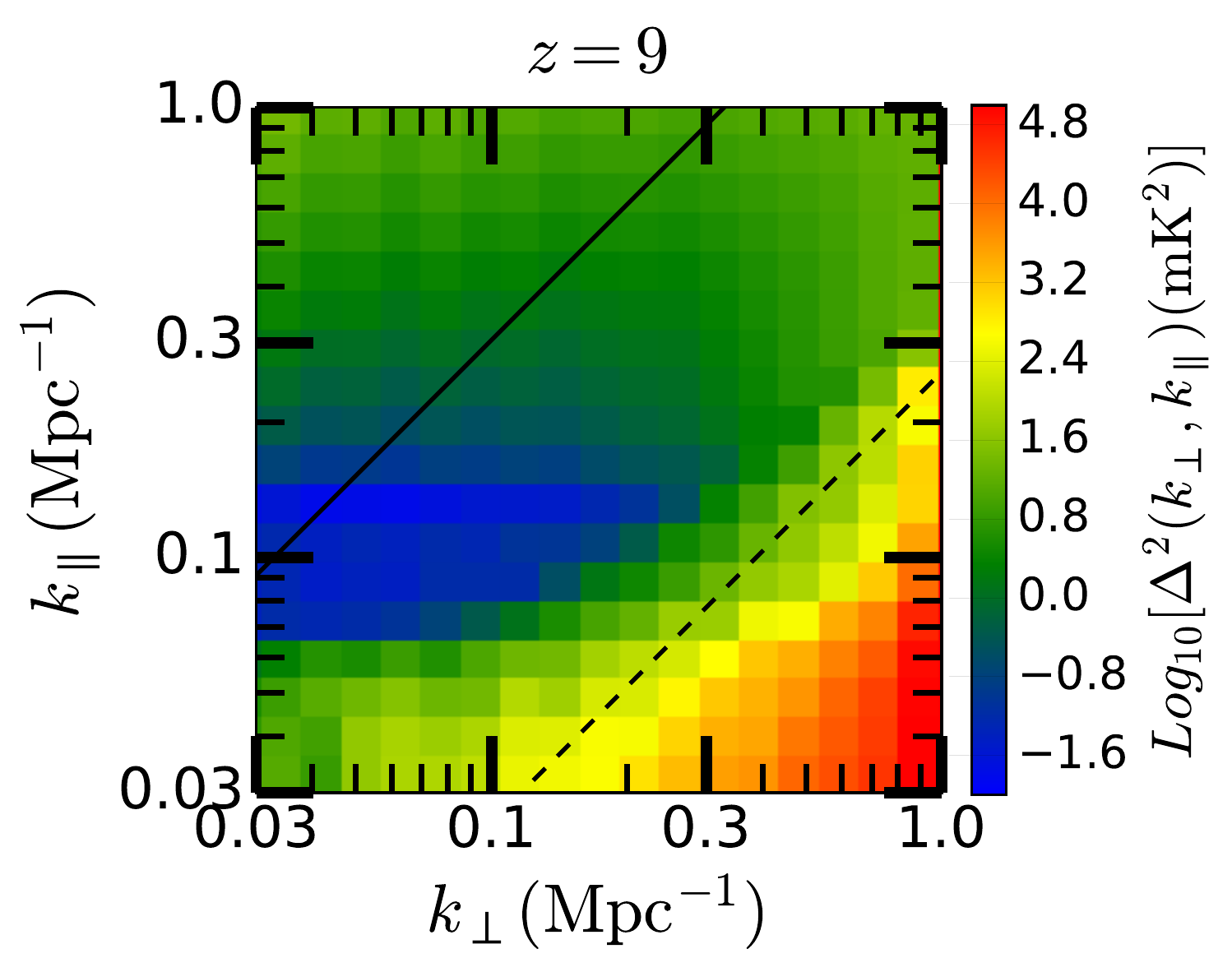}
\caption{Foreground noise in $k_{\perp}\mbox{-}k_{\parallel}$ space for 10\%uncertainty in the spectral index of GDSE.}
\label{FGBeta}
\end{figure}
\begin{figure}
\includegraphics[width=0.5\textwidth]{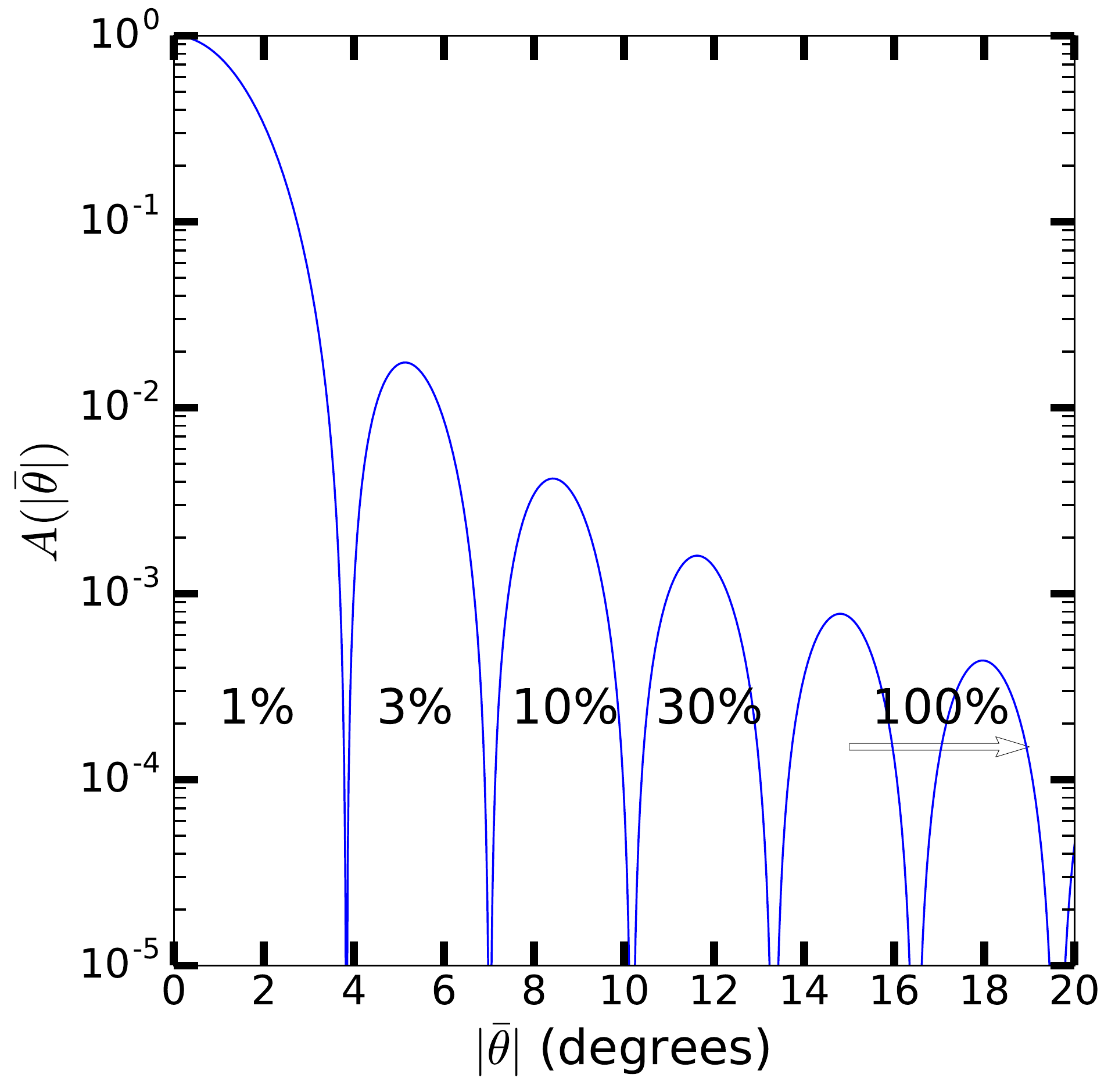}
\caption{Primary beam for doing simulations of Foreground Noise (Figures~\ref{FGAmp},\ref{FGBeta}). The numbers at the bottom indicate percentage by which 
the beam is assumed to be uncertain (Figure~\ref{FGBeam}).}
\label{Airy}
\end{figure}
\begin{figure}
\includegraphics[width=0.5\textwidth]{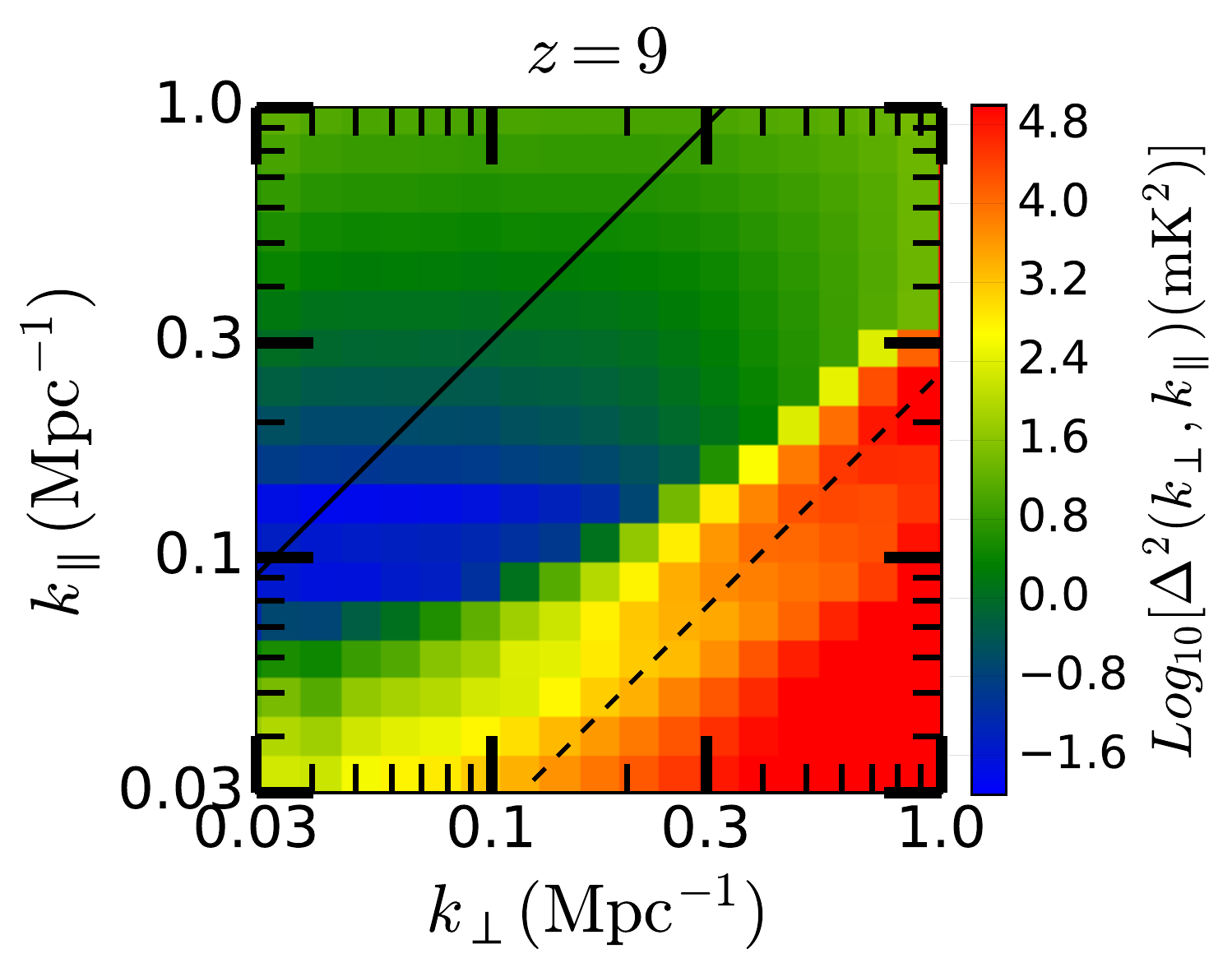}
\caption{Noise due to uncertainty in the determination of primary beam.}
\label{FGBeam}
\end{figure}

The Galactic Diffuse Synchrotron Emission (or GDSE) model considered in this paper is:
\begin{equation}
T_{GDSE} = A_{\mathrm{syn}} \left( \frac{\nu}{\nu_c} \right)^{\beta}
\label{eq:GDSE}
\end{equation}
$A_{\mathrm{syn}}$ is assumed to be 351K with 10\% variation in its value across different Lines of Sights. $\beta$, the
spectral index, is assumed to be about
2.55 (running spectral component is also there with value of about 0.1) \citep{2008AJ....136..641R}.
For the temperature distribution given by Equation~\ref{eq:GDSE}, one can simulate the visibilities \citep{2017isra.book.....T,1999ASPC..180.....T} 
and compute the power corresponding to these visibilities.
For computing visibilities corresponding to the Foregrounds considered, entire sky was pixelized in 625 pixels. 
The pixelization of the sky is denser at the centre of the primary beam and coarser towards the horizon.
The amplitudes of GDSE for each pixel
are chosen to be Gaussian variables with mean value of 351K and standard deviation of 10\% of the mean value. $\nu_c$ is 150 MHz.
Once one has the mean realization, aim is to deviate around this mean realization and compute the noise in power arising from 
uncertainties in the Foreground parameters.
As we know, the Fourier transformed visibilities are directly related to the power.
One can now obtain multiple realizations (about 10) of the same and then look for standard deviation in the Foreground power obtained for
different realizations. This would give rise to what I have been calling as Foreground Noise. Note that I have used extended Blackmann-Nuttall window \citep{2015ApJ...804...14T} for simulating Foreground visibilities along the frequency axis.
One can estimate the foreground noise if there is say 2\% uncertainty in the estimation of GDSE amplitude along each LOS. 
This is shown in Figure~\ref{FGAmp}.
One can also estimate the Foreground noise if the $\beta$ parameter is uncertain by 10\% along each LOS. This is shown in Figure~\ref{FGBeta}. 
As seen from the figures, the noise generated is comparable to the signal for the quoted uncertainties.
One can also get a rough estimate of the Foreground noise for the case of GDSE amplitude uncertainty.
If the primary beam falls
to about 1\% of its peak value for the first sidelobe. And if the error in the estimation of amplitude of GDSE is assumed to be about 1\%. 
Then, for power spectrum of Foregrounds, this would correspond to a roughly $(.01\times.01)^2=10^{-8}$ amount of error and so this (noise in) power 
is seen to be comparable to the signal power. The Foreground noise is much larger in the wedge region as the most effective part of the primary beam is 
larger in that region.
The beam for above analysis was assumed to be Airy pattern with first null occurring at about 4 degrees and the extent of the 
primary beam is all the way up-to the Horizon.
Note that this could be a pessimistic scenario as the beam
would fall more rapidly as one observes the sky through higher order side-lobes.
 One can also check
the effect of beam calibration errors on the measurement of power spectrum. The results are as depicted in Figure~\ref{FGBeam}.
Here foreground parameters are fixed while the beam is assumed to have some percentage variations about its mean value. The extents
 of variations are as depicted in Figure~\ref{Airy}. 

As seen from these figures, the region between the Foreground wedge and FoV wedge has small noise and hence could yield reasonable measurements of the EoR signal.

\section{Discussion}

The main conclusion of this paper is that one could do the FG modelling, FG subtraction and signal estimation analysis in the 
most of the Fourier space that is inclusive of the region in between the FoV wedge and the Horizon wedge. 
This is also the $k$-space with mapping of one space to another as explained in section~\ref{sec:theory}.
The disadvantage of working in the Fourier space is that there is some loss of signal power, say about 5-10\%, especially at
larger baselines. The signal can still remain above the thermal noise and foreground noise, provided one samples for large durations.
The main advantage of working in the Fourier space is that we directly deal with visibilities. We do not have to do a
deconvolution with primary beam to get image cube or temperature maps. 
The procedure
of going from visibilities to temperature maps is not at all trivial as we are dealing with foregrounds which are about $10^4 \mbox{-} 10^5$
times bigger than the signal. Also, as mentioned in the introduction, the foregrounds would have to be measured to an accuracy of
one part in $10^4 \mbox{-} 10^5$ to facilitate utilization of the entire Fourier space.
In this paper, I have proposed a simple method of dealing with foregrounds. One could average visibilities over a wide 
frequency channels and determine foreground parameters from these visibilities. Signal, being rapidly fluctuating with respect to 
frequency over frequency range $\ge 5MHz$, would tend to average out to a very small value. If one fits physically motivated model 
of foregrounds to the images
obtained from coarse-grained visibilities then one is expected to achieve good handle on foregrounds along with errors on foreground
parameters. These foreground model can be propagated through the instrument pipeline to simulate the visibilities and also uncertainties
in them in the Fourier space. These uncertainties would act as foreground noise and can be added in quadrature to the thermal noise.
Thus once one subtracts foreground visibilities, one gets signal visibilities with uncertainties. Writing technically:
\begin{equation}
V_{\rm{Total}}({\mathbf U,\eta}) = V_{\rm{Signal}}({\mathbf U,\eta}) + V_{\rm{FG}}({\mathbf U,\eta}) + V_{\rm{Other}}({\mathbf U,\eta})
\end{equation}
Here, $V_{\rm{Other}}({\mathbf U,\eta})$ includes other sources of visibilities like instrumental mode-mixing.  
Contribution of the this term to total visibilities could to be modelled. Thus its power contribution could be estimated and subtracted out. It would also carry noise and that should be added to other noise contributors.
One would have to model the chromatic response of the instrument, \citep[see e.g.][]{2013ApJ...770..156H,2015ApJ...804...14T} and also the calibrations and mis-direction errors because of point sources \citep[see e.g.][]{2010ApJ...724..526D,2012ApJ...752..137M}. 
While the scintillation noise could be added to the other sources of noise \citep[see e.g.][]{2010ApJ...718..963K,2016MNRAS.458.3099V}.
\par
\begin{figure}
\includegraphics[width=0.5\textwidth]{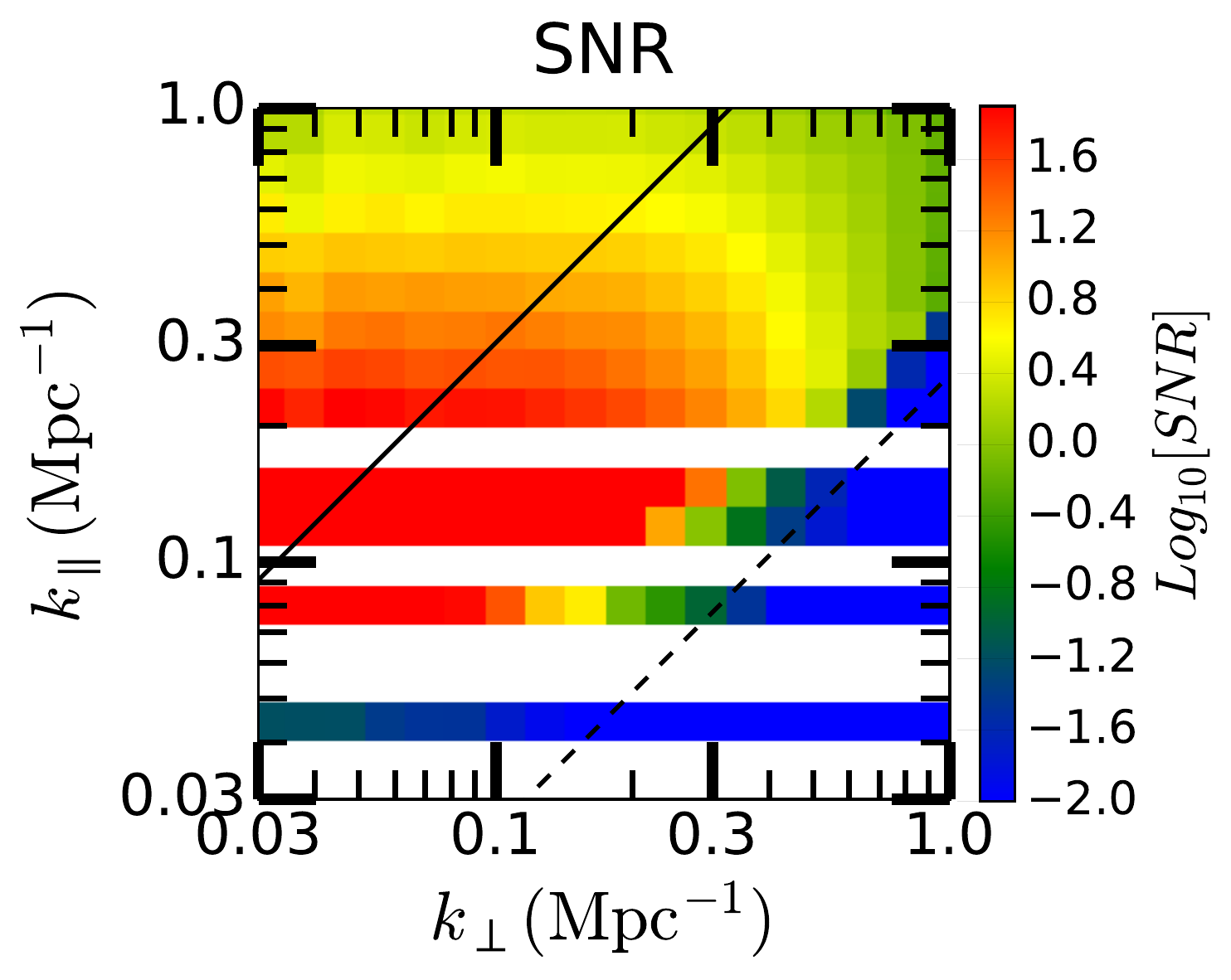}
\caption{Signal to Noise Ratio. Uncertainty in the signal is coming from FG noise due to errors in amplitude of GDSE only
and thermal noise.}
\label{SNR}
\end{figure}

Note that in the above equation, the quantity on the left hand side has intrinsic standard deviation of the thermal noise. Now,
\begin{equation}
V_{\rm{Signal}}({\mathbf U,\eta}) = V_{\rm{Total}}({\mathbf U,\eta}) - V_{\rm{FG}}({\mathbf U,\eta}) - V_{\rm{Other}}({\mathbf U,\eta})
\end{equation}
Thus for the signal visibilities one gets
\begin{equation}
\sigma_{V_{\rm{Signal}}({\mathbf U,\eta})} = \sqrt{ \sigma^2_{V_{\rm{Thermal}}({\mathbf U,\eta})} + \sigma^2_{V_{\rm{FG}}({\mathbf U,\eta})} + 
\sigma^2_{V_{\rm{Other}}({\mathbf U,\eta})} }
\end{equation}
Once one subtracts the foreground component this total deviation would be the one on signal and would yield total uncertainty in
the signal. The Signal to Noise Ratio (SNR) for FG noise due to errors in the GDSE amplitude and thermal noise, added in quadrature, is plotted in Figure~\ref{SNR}. As seen from the plot, there is considerable amount of region in between the two lines that can yield
significant information about the signal from EoR.
\par
There would be many sources that would contribute to the $\sigma_{V_{\rm{Signal}}({\mathbf U,\eta})}$ and they are not limited to
parameters of the foreground. Even calibration errors in the primary beam or uncertainties in the positions of bright point sources
would  contribute to the $\sigma_{V_{\rm{Signal}}({\mathbf U,\eta})}$ and would add to total noise.
Thus the SNR shown in Figure~\ref{SNR} is an estimate and as mentioned earlier it does not include noise generate due to 
instrumental mode mixing or other possible sources of errors.
These extra sources of errors like ionosphere, direction dependent beams, bandpass, etc. 
pose significant challenges as they form excess contaminants and make the removal of the smooth foreground component very difficult.
 A careful analysis of all the
effects would need much more effort and hence it has been deferred as future work. There one can address the question as to what
sort of calibration requirements, both instrument and foreground, are needed so that the total noise remains controlled and below the signal
strength. 

\section*{Appendix}
We have, 
\begin{equation}
V({\mathbf U},\eta)
= \int_{\Omega,B} d^2 \theta d\nu A_{\nu}({\pmb{\theta}}) I_{\nu}({\pmb {\theta}})  e^{- 2 \pi i \left( {\mathbf U} \cdot {\pmb{\theta}} + \nu \eta \right)} 
\end{equation}
Now substituting as follows: ($D_c(z)$ is comoving distance to redshift $z$):
\begin{eqnarray*}
\pmb{\theta} &=& \mathbf{r}_{\perp}/D_c(z) \\
\mathbf{U} &=& \mathbf{d}/\lambda = \mathbf{d} \nu/c
\end{eqnarray*}
and writing $\nu$ in terms of LOS distance $r_{\parallel}$ ($\nu_0$ is the central frequency and $r'_{\parallel}$ is derivative of LOS distance
with respect to frequency),
\begin{eqnarray*}
d\nu &=& dr_{\parallel} / r'_{\parallel} \\
\nu &=& \nu_0 + \Delta \nu = \nu_0 + \Delta r_{\parallel} /  r'_{\parallel}
\end{eqnarray*}
yields,
\begin{eqnarray*}
V({\mathbf U},\eta)
&=& \frac{1}{D_c(z)^2 r'_{\parallel}} \int d^2 {\mathbf{r}}_{\perp} dr_{\parallel} A({\mathbf{r}}_{\perp},r_{\parallel}) I({\mathbf{r}}_{\perp},r_{\parallel}) \\
&\times& e^{- 2 \pi i \left( {\mathbf d} [\nu_0 + \Delta r_{\parallel}/r'_{\parallel}]/c \cdot {{\mathbf{r}}_{\perp}/D_c(z)} + [\nu_0+ \Delta r_{\parallel}/r'_{\parallel}] \eta \right)} 
\end{eqnarray*}
As we use periodic boundary conditions in cosmology analysis, the origin can be shifted to the centre of the observation volume.
Thus $\Delta r_{\parallel}$ can be replaced by $r_{\parallel}$. 
Considering this, dropping the phase term $e^{-2 \pi i \nu_0 \eta}$ (as an overall phase would not matter for squared visibilities) and identifying 
\begin{eqnarray*}
{\mathbf k}_{\perp} &=& \frac{2 \pi}{D_c(z)} {\mathbf d} \nu_0/c \\
k_{\parallel} &=& \frac{2 \pi}{r'_{\parallel}} \eta 
\end{eqnarray*}
one gets the desired equation, 
\begin{eqnarray*}
V({\mathbf k}_{\perp},k_{\parallel}) &=& \frac{1}{r'_{\parallel}D_c(z)^2} \int d r_{\parallel} d^2r_{\perp} A(r_{\parallel},{\mathbf r}_{\perp})
I(r_{\parallel},{\mathbf r}_{\perp}) \\ 
&\times&
\exp \left[-i \left ( {\mathbf r}_{\perp} \cdot {\mathbf k}_{\perp} + r_{\parallel} k_{\parallel} \right) \right] \exp \left( - 2 \pi i \frac{{\mathbf r}_{\perp} \cdot {\mathbf d}}{D_c(z) c} \frac{r_{\parallel}}{r'_{\parallel}} \right)
\end{eqnarray*}

\bibliography{main}{}

\begin{thebibliography}{}
\expandafter\ifx\csname natexlab\endcsname\relax\def\natexlab#1{#1}\fi
\providecommand{\url}[1]{\href{#1}{#1}}
\providecommand{\dodoi}[1]{doi:~\href{http://doi.org/#1}{\nolinkurl{#1}}}
\providecommand{\doeprint}[1]{\href{http://ascl.net/#1}{\nolinkurl{http://ascl.net/#1}}}
\providecommand{\doarXiv}[1]{\href{https://arxiv.org/abs/#1}{\nolinkurl{https://arxiv.org/abs/#1}}}

\bibitem[{{Ali} {et~al.}(2008){Ali}, {Bharadwaj}, \&
  {Chengalur}}]{2008MNRAS.385.2166A}
{Ali}, S.~S., {Bharadwaj}, S., \& {Chengalur}, J.~N. 2008, \mnras, 385, 2166,
  \dodoi{10.1111/j.1365-2966.2008.12984.x}

\bibitem[{{Ali} {et~al.}(2015){Ali}, {Parsons}, {Zheng}, {Pober}, {Liu},
  {Aguirre}, {Bradley}, {Bernardi}, {Carilli}, {Cheng}, {DeBoer}, {Dexter},
  {Grobbelaar}, {Horrell}, {Jacobs}, {Klima}, {MacMahon}, {Maree}, {Moore},
  {Razavi}, {Stefan}, {Walbrugh}, \& {Walker}}]{2015ApJ...809...61A}
{Ali}, Z.~S., {Parsons}, A.~R., {Zheng}, H., {et~al.} 2015, \apj, 809, 61,
  \dodoi{10.1088/0004-637X/809/1/61}

\bibitem[{{Beardsley} {et~al.}(2016){Beardsley}, {Hazelton}, {Sullivan},
  {Carroll}, {Barry}, {Rahimi}, {Pindor}, {Trott}, {Line}, {Jacobs}, {Morales},
  {Pober}, {Bernardi}, {Bowman}, {Busch}, {Briggs}, {Cappallo}, {Corey}, {de
  Oliveira-Costa}, {Dillon}, {Emrich}, {Ewall-Wice}, {Feng}, {Gaensler},
  {Goeke}, {Greenhill}, {Hewitt}, {Hurley-Walker}, {Johnston-Hollitt},
  {Kaplan}, {Kasper}, {Kim}, {Kratzenberg}, {Lenc}, {Loeb}, {Lonsdale},
  {Lynch}, {McKinley}, {McWhirter}, {Mitchell}, {Morgan}, {Neben},
  {Thyagarajan}, {Oberoi}, {Offringa}, {Ord}, {Paul}, {Prabu}, {Procopio},
  {Riding}, {Rogers}, {Roshi}, {Udaya Shankar}, {Sethi}, {Srivani},
  {Subrahmanyan}, {Tegmark}, {Tingay}, {Waterson}, {Wayth}, {Webster},
  {Whitney}, {Williams}, {Williams}, {Wu}, \& {Wyithe}}]{2016ApJ...833..102B}
{Beardsley}, A.~P., {Hazelton}, B.~J., {Sullivan}, I.~S., {et~al.} 2016, \apj,
  833, 102, \dodoi{10.3847/1538-4357/833/1/102}

\bibitem[{{Bonaldi} \& {Brown}(2015)}]{2015MNRAS.447.1973B}
{Bonaldi}, A., \& {Brown}, M.~L. 2015, \mnras, 447, 1973,
  \dodoi{10.1093/mnras/stu2601}

\bibitem[{{Bowman} {et~al.}(2006){Bowman}, {Morales}, \&
  {Hewitt}}]{2006ApJ...638...20B}
{Bowman}, J.~D., {Morales}, M.~F., \& {Hewitt}, J.~N. 2006, \apj, 638, 20,
  \dodoi{10.1086/498703}

\bibitem[{{Bowman} {et~al.}(2018){Bowman}, {Rogers}, {Monsalve}, {Mozdzen}, \&
  {Mahesh}}]{2018Natur.555...67B}
{Bowman}, J.~D., {Rogers}, A.~E.~E., {Monsalve}, R.~A., {Mozdzen}, T.~J., \&
  {Mahesh}, N. 2018, \nat, 555, 67, \dodoi{10.1038/nature25792}

\bibitem[{{Chapman} {et~al.}(2012){Chapman}, {Abdalla}, {Harker}, {Jeli{\'c}},
  {Labropoulos}, {Zaroubi}, {Brentjens}, {de Bruyn}, \&
  {Koopmans}}]{2012MNRAS.423.2518C}
{Chapman}, E., {Abdalla}, F.~B., {Harker}, G., {et~al.} 2012, \mnras, 423,
  2518, \dodoi{10.1111/j.1365-2966.2012.21065.x}

\bibitem[{{Choudhury}(2009)}]{2009CSci...97..841C}
{Choudhury}, T.~R. 2009, Current Science, 97, 841.
\newblock \doarXiv{0904.4596}

\bibitem[{{Datta} {et~al.}(2010){Datta}, {Bowman}, \&
  {Carilli}}]{2010ApJ...724..526D}
{Datta}, A., {Bowman}, J.~D., \& {Carilli}, C.~L. 2010, \apj, 724, 526,
  \dodoi{10.1088/0004-637X/724/1/526}

\bibitem[{{Di Matteo} {et~al.}(2004){Di Matteo}, {Ciardi}, \&
  {Miniati}}]{2004MNRAS.355.1053D}
{Di Matteo}, T., {Ciardi}, B., \& {Miniati}, F. 2004, \mnras, 355, 1053,
  \dodoi{10.1111/j.1365-2966.2004.08443.x}

\bibitem[{{Di Matteo} {et~al.}(2002){Di Matteo}, {Perna}, {Abel}, \&
  {Rees}}]{2002ApJ...564..576D}
{Di Matteo}, T., {Perna}, R., {Abel}, T., \& {Rees}, M.~J. 2002, \apj, 564,
  576, \dodoi{10.1086/324293}

\bibitem[{{Fan} {et~al.}(2006){Fan}, {Carilli}, \&
  {Keating}}]{2006ARA&A..44..415F}
{Fan}, X., {Carilli}, C.~L., \& {Keating}, B. 2006, \araa, 44, 415,
  \dodoi{10.1146/annurev.astro.44.051905.092514}

\bibitem[{{Furlanetto} {et~al.}(2006){Furlanetto}, {Oh}, \&
  {Briggs}}]{2006PhR...433..181F}
{Furlanetto}, S.~R., {Oh}, S.~P., \& {Briggs}, F.~H. 2006, \physrep, 433, 181,
  \dodoi{10.1016/j.physrep.2006.08.002}

\bibitem[{{Gehlot} {et~al.}(2019){Gehlot}, {Mertens}, {Koopmans}, {Brentjens},
  {Zaroubi}, {Ciardi}, {Ghosh}, {Hatef}, {Iliev}, {Jeli{\'c}}, {}, {Kooistra},
  {Krause}, {Mellema}, {Mevius}, {Mitra}, {Offringa}, {Pandey}, {Sardarabadi},
  {Schaye}, {Silva}, {Vedantham}, \& {Yatawatta}}]{2019MNRAS.488.4271G}
{Gehlot}, B.~K., {Mertens}, F.~G., {Koopmans}, L.~V.~E., {et~al.} 2019, \mnras,
  488, 4271, \dodoi{10.1093/mnras/stz1937}

\bibitem[{{Ghosh} {et~al.}(2015){Ghosh}, {Koopmans}, {Chapman}, \&
  {Jeli{\'c}}}]{2015MNRAS.452.1587G}
{Ghosh}, A., {Koopmans}, L.~V.~E., {Chapman}, E., \& {Jeli{\'c}}, V. 2015,
  \mnras, 452, 1587, \dodoi{10.1093/mnras/stv1355}

\bibitem[{{Ghosh} {et~al.}(2018){Ghosh}, {Mertens}, \&
  {Koopmans}}]{2018MNRAS.474.4552G}
{Ghosh}, A., {Mertens}, F.~G., \& {Koopmans}, L.~V.~E. 2018, \mnras, 474, 4552,
  \dodoi{10.1093/mnras/stx2959}

\bibitem[{{Gleser} {et~al.}(2008){Gleser}, {Nusser}, \&
  {Benson}}]{2008MNRAS.391..383G}
{Gleser}, L., {Nusser}, A., \& {Benson}, A.~J. 2008, \mnras, 391, 383,
  \dodoi{10.1111/j.1365-2966.2008.13897.x}

\bibitem[{{Harker} {et~al.}(2009){Harker}, {Zaroubi}, {Bernardi}, {Brentjens},
  {de Bruyn}, {Ciardi}, {Jeli{\'c}}, {Koopmans}, {Labropoulos}, {Mellema},
  {Offringa}, {Pandey}, {Schaye}, {Thomas}, \&
  {Yatawatta}}]{2009MNRAS.397.1138H}
{Harker}, G., {Zaroubi}, S., {Bernardi}, G., {et~al.} 2009, \mnras, 397, 1138,
  \dodoi{10.1111/j.1365-2966.2009.15081.x}

\bibitem[{{Harker} {et~al.}(2010){Harker}, {Zaroubi}, {Bernardi}, {Brentjens},
  {de Bruyn}, {Ciardi}, {Jeli{\'c}}, {Koopmans}, {Labropoulos}, {Mellema},
  {Offringa}, {Pandey}, {Pawlik}, {Schaye}, {Thomas}, \&
  {Yatawatta}}]{2010MNRAS.405.2492H}
---. 2010, \mnras, 405, 2492, \dodoi{10.1111/j.1365-2966.2010.16628.x}

\bibitem[{{Hazelton} {et~al.}(2013){Hazelton}, {Morales}, \&
  {Sullivan}}]{2013ApJ...770..156H}
{Hazelton}, B.~J., {Morales}, M.~F., \& {Sullivan}, I.~S. 2013, \apj, 770, 156,
  \dodoi{10.1088/0004-637X/770/2/156}

\bibitem[{{Jeli{\'c}} {et~al.}(2010){Jeli{\'c}}, {Zaroubi}, {Labropoulos},
  {Bernardi}, {de Bruyn}, \& {Koopmans}}]{2010MNRAS.409.1647J}
{Jeli{\'c}}, V., {Zaroubi}, S., {Labropoulos}, P., {et~al.} 2010, \mnras, 409,
  1647, \dodoi{10.1111/j.1365-2966.2010.17407.x}

\bibitem[{{Koopmans}(2010)}]{2010ApJ...718..963K}
{Koopmans}, L.~V.~E. 2010, \apj, 718, 963, \dodoi{10.1088/0004-637X/718/2/963}

\bibitem[{{Liu} {et~al.}(2014{\natexlab{a}}){Liu}, {Parsons}, \&
  {Trott}}]{2014PhRvD..90b3018L}
{Liu}, A., {Parsons}, A.~R., \& {Trott}, C.~M. 2014{\natexlab{a}}, \prd, 90,
  023018, \dodoi{10.1103/PhysRevD.90.023018}

\bibitem[{{Liu} {et~al.}(2014{\natexlab{b}}){Liu}, {Parsons}, \&
  {Trott}}]{2014PhRvD..90b3019L}
---. 2014{\natexlab{b}}, \prd, 90, 023019, \dodoi{10.1103/PhysRevD.90.023019}

\bibitem[{{Liu} \& {Tegmark}(2011)}]{2011PhRvD..83j3006L}
{Liu}, A., \& {Tegmark}, M. 2011, \prd, 83, 103006,
  \dodoi{10.1103/PhysRevD.83.103006}

\bibitem[{{Liu} {et~al.}(2009{\natexlab{a}}){Liu}, {Tegmark}, {Bowman},
  {Hewitt}, \& {Zaldarriaga}}]{2009MNRAS.398..401L}
{Liu}, A., {Tegmark}, M., {Bowman}, J., {Hewitt}, J., \& {Zaldarriaga}, M.
  2009{\natexlab{a}}, \mnras, 398, 401,
  \dodoi{10.1111/j.1365-2966.2009.15156.x}

\bibitem[{{Liu} {et~al.}(2009{\natexlab{b}}){Liu}, {Tegmark}, \&
  {Zaldarriaga}}]{2009MNRAS.394.1575L}
{Liu}, A., {Tegmark}, M., \& {Zaldarriaga}, M. 2009{\natexlab{b}}, \mnras, 394,
  1575, \dodoi{10.1111/j.1365-2966.2009.14426.x}

\bibitem[{{McQuinn} {et~al.}(2006){McQuinn}, {Zahn}, {Zaldarriaga},
  {Hernquist}, \& {Furlanetto}}]{2006ApJ...653..815M}
{McQuinn}, M., {Zahn}, O., {Zaldarriaga}, M., {Hernquist}, L., \& {Furlanetto},
  S.~R. 2006, \apj, 653, 815, \dodoi{10.1086/505167}

\bibitem[{{Mertens} {et~al.}(2018){Mertens}, {Ghosh}, \&
  {Koopmans}}]{2018MNRAS.478.3640M}
{Mertens}, F.~G., {Ghosh}, A., \& {Koopmans}, L.~V.~E. 2018, \mnras, 478, 3640,
  \dodoi{10.1093/mnras/sty1207}

\bibitem[{{Mesinger} {et~al.}(2011){Mesinger}, {Furlanetto}, \&
  {Cen}}]{2011MNRAS.411..955M}
{Mesinger}, A., {Furlanetto}, S., \& {Cen}, R. 2011, \mnras, 411, 955,
  \dodoi{10.1111/j.1365-2966.2010.17731.x}

\bibitem[{{Morales} {et~al.}(2019){Morales}, {Beardsley}, {Pober}, {Barry},
  {Hazelton}, {Jacobs}, \& {Sullivan}}]{2019MNRAS.483.2207M}
{Morales}, M.~F., {Beardsley}, A., {Pober}, J., {et~al.} 2019, \mnras, 483,
  2207, \dodoi{10.1093/mnras/sty2844}

\bibitem[{{Morales} {et~al.}(2012){Morales}, {Hazelton}, {Sullivan}, \&
  {Beardsley}}]{2012ApJ...752..137M}
{Morales}, M.~F., {Hazelton}, B., {Sullivan}, I., \& {Beardsley}, A. 2012,
  \apj, 752, 137, \dodoi{10.1088/0004-637X/752/2/137}

\bibitem[{{Morales} \& {Hewitt}(2004)}]{2004ApJ...615....7M}
{Morales}, M.~F., \& {Hewitt}, J. 2004, \apj, 615, 7, \dodoi{10.1086/424437}

\bibitem[{{Oh} \& {Mack}(2003)}]{2003MNRAS.346..871O}
{Oh}, S.~P., \& {Mack}, K.~J. 2003, \mnras, 346, 871,
  \dodoi{10.1111/j.1365-2966.2003.07133.x}

\bibitem[{{Parsons} {et~al.}(2012{\natexlab{a}}){Parsons}, {Pober}, {McQuinn},
  {Jacobs}, \& {Aguirre}}]{2012ApJ...753...81P}
{Parsons}, A., {Pober}, J., {McQuinn}, M., {Jacobs}, D., \& {Aguirre}, J.
  2012{\natexlab{a}}, \apj, 753, 81, \dodoi{10.1088/0004-637X/753/1/81}

\bibitem[{{Parsons} {et~al.}(2016){Parsons}, {Liu}, {Ali}, \&
  {Cheng}}]{2016ApJ...820...51P}
{Parsons}, A.~R., {Liu}, A., {Ali}, Z.~S., \& {Cheng}, C. 2016, \apj, 820, 51,
  \dodoi{10.3847/0004-637X/820/1/51}

\bibitem[{{Parsons} {et~al.}(2012{\natexlab{b}}){Parsons}, {Pober}, {Aguirre},
  {Carilli}, {Jacobs}, \& {Moore}}]{2012ApJ...756..165P}
{Parsons}, A.~R., {Pober}, J.~C., {Aguirre}, J.~E., {et~al.}
  2012{\natexlab{b}}, \apj, 756, 165, \dodoi{10.1088/0004-637X/756/2/165}

\bibitem[{{Patil} {et~al.}(2017){Patil}, {Yatawatta}, {Koopmans}, {de Bruyn},
  {Brentjens}, {Zaroubi}, {Asad}, {Hatef}, {Jeli{\'c}}, {Mevius}, {Offringa},
  {Pandey}, {Vedantham}, {Abdalla}, {Brouw}, {Chapman}, {Ciardi}, {Gehlot},
  {Ghosh}, {Harker}, {Iliev}, {Kakiichi}, {Majumdar}, {Mellema}, {Silva},
  {Schaye}, {Vrbanec}, \& {Wijnholds}}]{2017ApJ...838...65P}
{Patil}, A.~H., {Yatawatta}, S., {Koopmans}, L.~V.~E., {et~al.} 2017, \apj,
  838, 65, \dodoi{10.3847/1538-4357/aa63e7}

\bibitem[{{Paul} {et~al.}(2016){Paul}, {Sethi}, {Morales}, {Dwarkanath}, {Udaya
  Shankar}, {Subrahmanyan}, {Barry}, {Beardsley}, {Bowman}, {Briggs},
  {Carroll}, {de Oliveira-Costa}, {Dillon}, {Ewall-Wice}, {Feng}, {Greenhill},
  {Gaensler}, {Hazelton}, {Hewitt}, {Hurley-Walker}, {Jacobs}, {Kim},
  {Kittiwisit}, {Lenc}, {Line}, {Loeb}, {McKinley}, {Mitchell}, {Neben},
  {Offringa}, {Pindor}, {Pober}, {Procopio}, {Riding}, {Sullivan}, {Tegmark},
  {Thyagarajan}, {Tingay}, {Trott}, {Wayth}, {Webster}, {Wyithe}, {Cappallo},
  {Johnston-Hollitt}, {Kaplan}, {Lonsdale}, {McWhirter}, {Morgan}, {Oberoi},
  {Ord}, {Prabu}, {Srivani}, {Williams}, \& {Williams}}]{2016ApJ...833..213P}
{Paul}, S., {Sethi}, S.~K., {Morales}, M.~F., {et~al.} 2016, \apj, 833, 213,
  \dodoi{10.3847/1538-4357/833/2/213}

\bibitem[{{Petrovic} \& {Oh}(2011)}]{2011MNRAS.413.2103P}
{Petrovic}, N., \& {Oh}, S.~P. 2011, \mnras, 413, 2103,
  \dodoi{10.1111/j.1365-2966.2011.18276.x}

\bibitem[{{Planck Collaboration} {et~al.}(2016){Planck Collaboration}, {Ade},
  {Aghanim}, {Arnaud}, {Ashdown}, {Aumont}, {Baccigalupi}, {Banday},
  {Barreiro}, {Bartlett}, \& et~al.}]{2016A&A...594A..13P}
{Planck Collaboration}, {Ade}, P.~A.~R., {Aghanim}, N., {et~al.} 2016, \aap,
  594, A13, \dodoi{10.1051/0004-6361/201525830}

\bibitem[{{Pober} {et~al.}(2013){Pober}, {Parsons}, {Aguirre}, {Ali},
  {Bradley}, {Carilli}, {DeBoer}, {Dexter}, {Gugliucci}, {Jacobs}, {Klima},
  {MacMahon}, {Manley}, {Moore}, {Stefan}, \& {Walbrugh}}]{2013ApJ...768L..36P}
{Pober}, J.~C., {Parsons}, A.~R., {Aguirre}, J.~E., {et~al.} 2013, \apjl, 768,
  L36, \dodoi{10.1088/2041-8205/768/2/L36}

\bibitem[{{Pritchard} \& {Loeb}(2012)}]{2012RPPh...75h6901P}
{Pritchard}, J.~R., \& {Loeb}, A. 2012, Reports on Progress in Physics, 75,
  086901, \dodoi{10.1088/0034-4885/75/8/086901}

\bibitem[{{Rogers} \& {Bowman}(2008)}]{2008AJ....136..641R}
{Rogers}, A.~E.~E., \& {Bowman}, J.~D. 2008, \aj, 136, 641,
  \dodoi{10.1088/0004-6256/136/2/641}

\bibitem[{{Santos} {et~al.}(2005){Santos}, {Cooray}, \&
  {Knox}}]{2005ApJ...625..575S}
{Santos}, M.~G., {Cooray}, A., \& {Knox}, L. 2005, \apj, 625, 575,
  \dodoi{10.1086/429857}

\bibitem[{{Taylor} {et~al.}(1999){Taylor}, {Carilli}, \&
  {Perley}}]{1999ASPC..180.....T}
{Taylor}, G.~B., {Carilli}, C.~L., \& {Perley}, R.~A., eds. 1999, Astronomical
  Society of the Pacific Conference Series, Vol. 180, {Synthesis Imaging in
  Radio Astronomy II}

\bibitem[{{Thompson} {et~al.}(2017){Thompson}, {Moran}, \&
  {Swenson}}]{2017isra.book.....T}
{Thompson}, A.~R., {Moran}, J.~M., \& {Swenson}, Jr., G.~W. 2017,
  {Interferometry and Synthesis in Radio Astronomy, 3rd Edition},
  \dodoi{10.1007/978-3-319-44431-4}

\bibitem[{{Thyagarajan} {et~al.}(2015){Thyagarajan}, {Jacobs}, {Bowman},
  {Barry}, {Beardsley}, {Bernardi}, {Briggs}, {Cappallo}, {Carroll}, {Corey},
  {de Oliveira-Costa}, {Dillon}, {Emrich}, {Ewall-Wice}, {Feng}, {Goeke},
  {Greenhill}, {Hazelton}, {Hewitt}, {Hurley-Walker}, {Johnston-Hollitt},
  {Kaplan}, {Kasper}, {Kim}, {Kittiwisit}, {Kratzenberg}, {Lenc}, {Line},
  {Loeb}, {Lonsdale}, {Lynch}, {McKinley}, {McWhirter}, {Mitchell}, {Morales},
  {Morgan}, {Neben}, {Oberoi}, {Offringa}, {Ord}, {Paul}, {Pindor}, {Pober},
  {Prabu}, {Procopio}, {Riding}, {Rogers}, {Roshi}, {Udaya Shankar}, {Sethi},
  {Srivani}, {Subrahmanyan}, {Sullivan}, {Tegmark}, {Tingay}, {Trott},
  {Waterson}, {Wayth}, {Webster}, {Whitney}, {Williams}, {Williams}, {Wu}, \&
  {Wyithe}}]{2015ApJ...804...14T}
{Thyagarajan}, N., {Jacobs}, D.~C., {Bowman}, J.~D., {et~al.} 2015, \apj, 804,
  14, \dodoi{10.1088/0004-637X/804/1/14}

\bibitem[{{Trott} {et~al.}(2012){Trott}, {Wayth}, \&
  {Tingay}}]{2012ApJ...757..101T}
{Trott}, C.~M., {Wayth}, R.~B., \& {Tingay}, S.~J. 2012, \apj, 757, 101,
  \dodoi{10.1088/0004-637X/757/1/101}

\bibitem[{{Vedantham} {et~al.}(2012){Vedantham}, {Udaya Shankar}, \&
  {Subrahmanyan}}]{2012ApJ...745..176V}
{Vedantham}, H., {Udaya Shankar}, N., \& {Subrahmanyan}, R. 2012, \apj, 745,
  176, \dodoi{10.1088/0004-637X/745/2/176}

\bibitem[{{Vedantham} \& {Koopmans}(2016)}]{2016MNRAS.458.3099V}
{Vedantham}, H.~K., \& {Koopmans}, L.~V.~E. 2016, \mnras, 458, 3099,
  \dodoi{10.1093/mnras/stw443}

\bibitem[{{Wang} {et~al.}(2006){Wang}, {Tegmark}, {Santos}, \&
  {Knox}}]{2006ApJ...650..529W}
{Wang}, X., {Tegmark}, M., {Santos}, M.~G., \& {Knox}, L. 2006, \apj, 650, 529,
  \dodoi{10.1086/506597}

\end{thebibliography}
\bibliographystyle{aasjournal}
\end{document}